\documentclass[useAMS,usenatbib]{mnras}
\usepackage{graphicx} 
\usepackage{subcaption}
\usepackage{amsmath} 
\usepackage{xcolor}
\usepackage{float}
\usepackage{caption}
\usepackage{placeins}
\usepackage{lipsum}
\usepackage{multicol}
\usepackage{soul}
\usepackage{amssymb}
\usepackage{ragged2e}
\usepackage{threeparttable}
\usepackage{comment}
\def\refjnl#1{{\rm#1}}
\def\aj{\refjnl{AJ}}                   
\def\apj{\refjnl{ApJ}}                 
\def\apjl{\refjnl{ApJ}}                
\def\apjs{\refjnl{ApJS}}               
\def\ao{\refjnl{Appl.~Opt.}}           
\def\aap{\refjnl{A\&A}}                
\def\mnras{\refjnl{MNRAS}}             
\def\pasa{\refjnl{PASA}}               
\def\pasj{\refjnl{PASJ}}               
\def\nat{\refjnl{Nature}}              
\def\physrep{\refjnl{Phys.~Rep.}}   

\def\be{\begin{equation}} 
\def\ee{\end{equation}}

\def\msun{{\Msun}}

\def\gsim{\lower.5ex\hbox{\gtsima}} 
\def\lsim{\lower.5ex\hbox{\ltsima}} \def\gtsima{$\; \buildrel > \over 
\sim \;$} \def\ltsima{$\; \buildrel < \over \sim \;$} \def\prosima{$\; 
\buildrel \propto \over \sim \;$} \def\gsim{\lower.5ex\hbox{\gtsima}} 
\def\lsim{\lower.5ex\hbox{\ltsima}} 
\def\simgt{\lower.5ex\hbox{\gtsima}} 
\def\simlt{\lower.5ex\hbox{\ltsima}} 
\def\simpr{\lower.5ex\hbox{\prosima}} \def\la{\lsim} \def\ga{\gsim} 
  
 \def\gtsima{$\; \buildrel > \over \sim \;$} 
\def\ltsima{$\; \buildrel < \over \sim \;$} 
\def\gsim{\lower.5ex\hbox{\gtsima}} 
\def\lsim{\lower.5ex\hbox{\ltsima}} 
\def\simgt{\lower.5ex\hbox{\gtsima}} 
\def\simlt{\lower.5ex\hbox{\ltsima}} 
\def\simpr{\lower.5ex\hbox{\prosima}} 
\def\la{\lsim} 
\def\ga{\gsim}

\def\msun{\,{\rm \Msun}}

\def\E3{{\cal E}_{\rm g}^{III}}

\def\msun{\rm M_\odot}

\def\Msun{\rm M_\odot}

\def\M*{M_*}

\def\Z*{Z_*}
\def\L*{L_*}

\def \fesc{f_{\rm esc}}
\def \fescum {f_{\mathrm{esc}}^{\mathrm{cum}}}
\def \avfesc {\langle f_{\mathrm{esc}}\rangle}

\def\nho{n_{\mathrm{H}}^{0}}
\def\effesc{\langle \fesc \rangle_{\rm L_{ion}}}

\title[Modelling $f_{esc}$ in the EoR]{Modelling the escape of Lyman Continuum photons from galaxies in the Epoch of Reionization}
\author[Bremer \& Dayal]{Jonas Bremer$^{1}$\thanks{bremer@astro.rug.nl} and Pratika Dayal$^1$\\ 
$^{{1}}$ Kapteyn Astronomical Institute, University of Groningen, PO Box 800, 9700 AV Groningen, The Netherlands\\}

\begin{document}
\maketitle
\label{firstpage}
\pagerange{\pageref{firstpage}--\pageref{lastpage}}

\begin{abstract}
We couple the DELPHI framework for galaxy formation with a model for the escape of ionizing photons to study both its variability with galaxy assembly and the resulting key reionization sources. In this model, leakage either occurs through a fully ionized gas distribution (ionization bounded) or additionally through channels cleared of gas by supernova explosions (ionization bounded + holes). The escape fraction is therefore governed by a combination of the density and star formation rate. Having calibrated our star formation efficiencies to match high-$z$ observables, we find the central gas density to regulate the boundary between high ($\gsim 0.70$) and low ($\lsim 0.06$) escape fractions. As galaxies become denser at higher redshifts, this boundary shifts from $M_{h}\simeq 10^{9.5}\mathrm{M_{\odot}}$ at $z\sim 5$ to $M_{h}\simeq10^{7.8}\mathrm{M_{\odot}}$ at $z\sim 15$. While leakage is entirely governed through holes above this mass range, it is not affecting general trends for lower masses. We find the co-evolution of galaxy assembly and the degree of leakage to be mass and redshift dependent, driven by an increasing fraction of $f_{\mathrm{esc}}\lsim 0.06$ galaxies at increasing mass and redshift. The variability in the escape of ionizing photons is driven by the underlying variations in our dark matter assembly histories. Galaxies with $M_h \lsim 10^{7.9} ~ (10^{8.9})\msun$ provide half of the escaping ionizing emissivity by $z \sim 10 ~ (5)$ in the ionization bounded model. On the other hand, galaxies that purely leak through holes contribute $6$ $(13)\%$ at $z\sim 5$ $(15)$. Reionization ends slightly ($\sim 50\mathrm{Myr}$) earlier in the ionization bounded + holes model, leaving the overall shape of the reionization history unaffected. We end by exploring the impact of two reionization feedback scenarios, in which we suppress the gas content of galaxies with $T_{\mathrm{vir}}<20000\mathrm{K}$ and $v_{c}<30\mathrm{kms^{-1}}$ residing in ionized regions.
\end{abstract}


\section{Introduction}
The appearance of the first galaxies at $z\sim 20-30$ led to the production of Lyman Continuum (LyC; with energy $>13.6 \mathrm{ev}$) photons which gradually ionized the hydrogen in  the intergalactic medium \citep[IGM;][]{barkana2001, dayal2018} within the first billion years. This last major phase transition of all of the hydrogen in the IGM is termed the epoch of reionization (EoR). Analyses of high-redshift Quasar spectra indicate that hydrogen reionization nears completion by $z\sim 6$ \citep[e.g.][]{fan2006,becker2015,eilers2018, becker2021} and hint towards the EoR being a very patchy process \citep[e.g.][]{davies2016, eilers2018}. Additional constraints are derived from the Cosmic Microwave background (CMB) electron scattering optical depth \citep[e.g.][]{planck2018} and Lyman Alpha emitters \citep[LAEs; e.g.][]{stark2010,pentericci2011, curtislake2012,schenker2014, debarros2017}.

There is growing consensus that star-forming galaxies dominate the photon budget for reionization \citep[e.g.][]{robertson2015, dayal2020,naidu2020, trebitsch2022} with black-hole powered Active Galactic Nuclei (AGN) playing a minor role over the bulk of the EoR \citep[e.g.][]{becker2013,daloisio2017,mitra2018, kulkarni2019,dayal2020}. However, a key unknown involved in all such calculations concerns the ``escape fraction" of LyC photons that can escape into the IGM ($\fesc$).

Several efforts have been made to observationally constrain $\fesc$ through direct detection of the LyC emission at low to intermediate redshifts. Local to low redshift ($z \lsim 0.45$) observations yield values of $\fesc$ that range between $\sim 2-73\%$ \citep[][]{leitet2011,leitet2013,izotov2016a,izotov2016b,izotov2018a,izotov2018b,izotov2021}. Intermediate redshift observations ($z \sim 2.5-4$) yield a similarly wide range of value such that $\fesc \sim 15-60\%$ \citep[][]{shapley2016,vanzella2016,bian2017,vanzella2018, fletcher2019} although, as might be expected, stacking of non-detections up to $z\sim 3.5$ leads to lower upper limits of $\fesc \lsim 7\%$ \citep{ruthkowski2017,grazian2017,naidu2018, sexana2021}. Finally, 7 gravitationally lensed galaxies at $4<z<5$ have been used to infer $\fesc \simeq 19 \%$ \citep{leethochawalit2016}. Directly detecting LyC emission at higher-redshifts is unfeasible due the elevated opacity of an increasingly neutral IGM. Therefore at higher redshifts, $f_{\mathrm{esc}}$ determinations rely on indirect methods such as nebular emission signatures \citep[e.g.][]{anders2003,zackrisson2013, nakajima2014, izotov2021} or the Lyman Alpha line profile \citep[e.g.][]{verhamme2015,verhamme2017, izotov2021}. 

The dependence of $\fesc$ on physical properties such as the stellar mass too remains a matter of debate: while some works indicate a possible decline of $\fesc$ with stellar mass \citep{fletcher2019,sexana2021}, others find no evidence for such a relation \citep{izotov2021}. Interpreting such trends however remains challenging in light of model dependencies, source-to-source variation, poor statistics, and IGM line of sight effects. The low success rate of LyC detections and the large scatter in the observed estimates hint at the importance of other effects such as directional leakage and time variability. 
A growing body of work has also focused on simulating $\fesc$ using a multitude of approaches. These include hydrodynamic simulations \citep{gnedin2008,wise2009,kimm2014,wise2014,ma2015,paardekooper2015,xu2016,kimm2017,trebitsch2017,lewis2020}, constraining the global averaged escape fraction by matching the observed reionization history using analytically inclined approaches \citep[e.g.][]{inoue2006, robertson2013,robertson2015, mitra2015, sharma2016, dayal2017, dayal2020,naidu2020} or using analytic models \citep{fernandez2011,benson2013, ferarra2013}. While most of these works find $\fesc$ to decline with halo mass, a number find the opposite trend \citep{gnedin2008,sharma2016,naidu2020}. 

Simulating $\fesc$ has remained a particularly challenging issue because of complexities such as the fact that the bulk of LyC photons are absorbed in the vicinity of the newly born stars at the molecular cloud scale \citep{ma2015,paardekooper2015,kimm2017,trebitsch2017} and the poorly understood role of binary stars \citep[e.g][]{ma2016}, runaway OB stars \citep[e.g][]{kimm2014,ma2015} and the initial mass function \citep[IMF; e.g][]{wise2009}. The exact leakage mechanism (e.g. the role of Supernova versus photoionization feedback) too remains debated \citep{trebitsch2017, kimm2017}. Indeed, if low-density channels created by Supernovae (SN) explosions are the key escape mechanism, one must account for the time delay between ionizing photon production and escape \citep[e.g.][]{ferarra2013, ma2015, kimm2017}. Finally, the stochastic nature of the star formation process and its co-evolution with the structure of the interstellar medium (ISM) leads to an escape fraction that can fluctuate between 0 and 100\% on $\sim 10 ~\mathrm{Myr}$ time scales \citep{wise2009,wise2014, ma2015,kimm2017,trebitsch2017} and is highly non-isotropic \citep{gnedin2008,wise2009, wise2014, ma2015, paardekooper2015, trebitsch2017}, with galaxies with larger escape fractions leaking more extensively on larger angular scales \citep[e.g][]{paardekooper2015}. 

This work aims to address issues including: (i) the dependence of $\fesc$ on intrinsic galaxy properties; (ii) the time evolution of leakage concurrent with galaxy assembly; and (iii) understanding the key sources responsible for reionization. To answer these questions, we combine the semi-analytical framework {\sc Delphi} for high-$z$ ($z>4.5$) galaxy formation, with an analytic model for the evolution of ionization fronts within the ISM of galaxies \citep{ferarra2013}. The strength of this semi-analytic approach lies in that it enables us to study the coupling between high-redshift galaxy assembly and the associated escape fraction (and its time evolution) across a large dynamic range in mass. 

In this work, we use the following cosmological parameters ($\Omega_{\rm m }, \Omega_{\Lambda}, \Omega_{\rm b}, h, n_s, \sigma_8) =$ \citep[0.3111, 0.6889, 0.049, 0.68, 0.97, 0.81][]{planck2018} where $\Omega_{\rm m },\Omega_{\Lambda}, \Omega_{\rm b}$ indicate the cosmological density parameters for matter, Dark Energy and baryons, respectively, $h$ is the Hubble value, $n_s$ is the spectral index of the initial density perturbations and $\sigma_8$ represents the root mean square of density fluctuations on scales of $8 h^{-1}$ cMpc. All quantities are expressed in comoving units unless mentioned otherwise. 

This paper is structured as follows: we start by explaining our theoretical model in Sec. \ref{modelsection}, followed by Sec. \ref{fescsection} in which we address $f_{\mathrm{esc}}$ trends with galaxy properties, its time evolution and its variability. The implications of these results in the context of reionization are discussed in Sec. \ref{riosection} and we conclude in Sec. \ref{conclusionsection}.

\section{Theoretical Model}
\label{modelsection}
In this work we use the semi-analytical framework {\sc Delphi} \citep[\textbf{D}ark  Matter  and  the \textbf{e}mergence  of  ga\textbf{l}axies  in  the  e\textbf{p}oc\textbf{h} of  re\textbf{i}onization;][]{dayal2014}, designed to jointly track the dark matter and baryonic assembly of high-redshift ($z \gsim 4.5$) galaxies. We couple this framework with a model that describes the leakage of hydrogen ionizing photons, as detailed in what follows.

\subsection{The Semi-analytic framework}
We briefly describe the {\sc Delphi} model whose complete details can be found in \citet{dayal2014}. We start by using the 
modified binary merger tree algorithm described in \citet{parkinson2008} to build merger trees for 550 galaxies at $z=4.5$ uniformly distributed in the halo mass range of ${\rm log}(M_h/ \Msun)=7.5-13$ with a resolution of $M_{\mathrm{res}} = 10^{7}M_{\odot}$. These merger trees are built up to $z \sim 40$ in equal time-steps of $30 {\rm Myr}$ so that all type II supernovae (SNII) from a given stellar population explode within a time-step \citep{padovani1993}. Each $z=4.5$ parent halo is assigned a co-moving number density by matching to the $z=4.5$ Sheth-Tormen halo mass function \citep[HMF;][]{sheth1999} and this number density is propagated throughout the merger tree of that halo. We have checked that the resulting HMFs are in agreement with the Sheth-Tormen one at all redshifts. 

In terms of the baryonic physics, the first progenitors (starting leaves) of any galaxy are assigned an initial gas mass $M_{g}^{i}(t) = (\Omega_\mathrm{b}/\Omega_\mathrm{m}) M_{h}(t)$; this assumes that these halos have a gas-to-dark matter fraction equal to the cosmological one. However, for galaxies that have progenitors, the initial gas mass is the sum of the gas mass smoothly accreted from the IGM and the final gas mass brought in by its merging progenitors (accounting for star formation and SN feedback). The accreted gas mass is computed assuming that accreted dark matter drags in a cosmological gas mass fraction.

\begin{figure*}
\center{\includegraphics[scale=0.37]{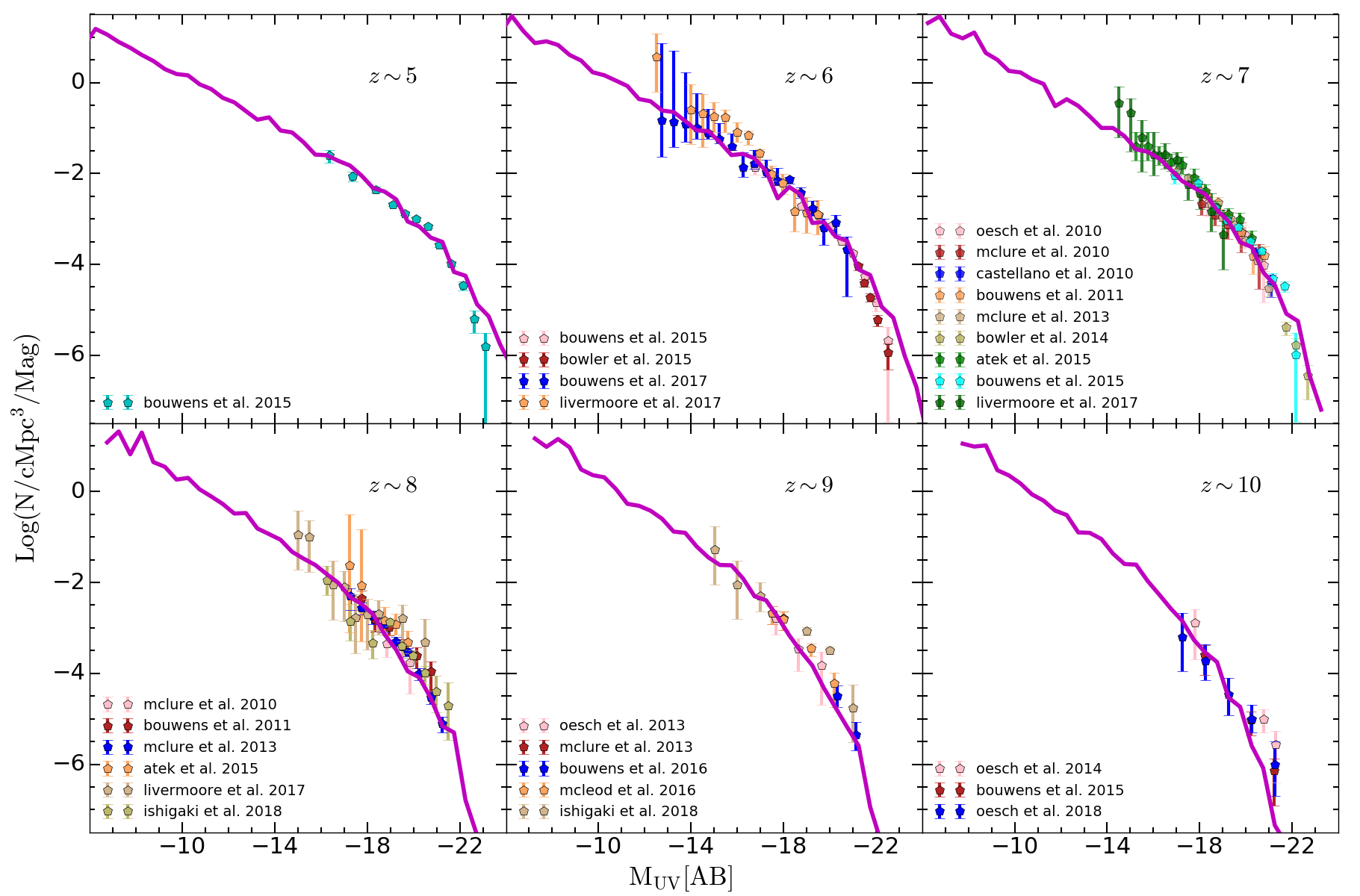}}
\caption{Calibrated UV Luminosity Function at $z\sim 5-10$ (magenta lines) for which we have used  $f_{*}=0.03$ and $f_{w}=0.10$. The data points represent observational estimates, collected from the following works \citep[][]{atek2015,bouwens2010a,bouwens2011b,bouwens2015,bouwens2016,bouwens2017,bowler2014a,bowler2015,castellano2010,ishigaki2017,livermore2017,mclure2010,mclure2013,mcleod2016,oesch2010,oesch2013,oesch2014}.}
\label{uvluminosityfunction}
\end{figure*}
The fraction of the initial gas mass transformed into stars is set by the effective star formation efficiency which is a minimum between the efficiency required to unbind the rest of the gas ($f_{*}^{\mathrm{ej}}$) up to a maximum threshold ($f_{*}$), i.e.
$f_{*}^{\mathrm{eff}}=\mathrm{min}[f_{*}^{\mathrm{ej}},f_{*}]$. Low-mass galaxies are feedback limited and form stars with an efficiency $f_{*}^{\mathrm{eff}}= f_{*}^{\mathrm{ej}}$ while  massive systems that are able to retain their gas mass form stars at $f_{*}^{\mathrm{eff}}= f_{*}$. As might be expected, $f_{*}^{\mathrm{ej}}$ depends on the fraction of SNII energy that can couple to the ISM gas ($f_w$). Our model therefore has only two mass- and redshift-independent free parameters ($f_w$ and $f_{*}$). 

\begin{figure}
\center{\includegraphics[width=\linewidth]{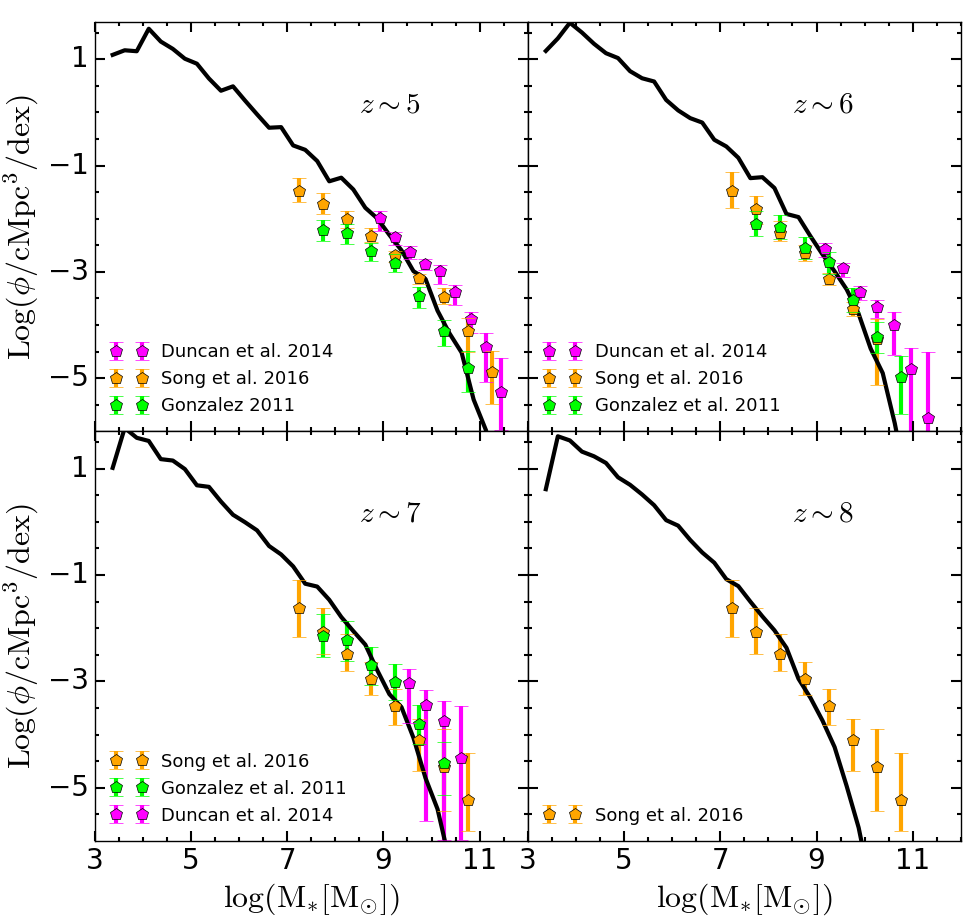}}
\caption{Calibrated Stellar Mass Function at $z\sim 5-8$ (magenta lines) for which we have used  $f_{*}=0.03$ and $f_{w}=0.10$. The data points represent observational estimates, collected from the following works \citep[][]{duncan2014,gonzalez2011,song2016}.}
\label{stellarmassfunction}
\end{figure}

Throughout this work we assume a Salpeter initial mass function \citep{salpeter1955} between $0.1-100\mathrm{M_{\odot}}$. We assume each new stellar population to have a metallicity of $Z=0.05Z_{\odot}$ and an age of $t_{0} =2 ~\mathrm{Myr}$ to compute its spectrum using the stellar population synthesis code STARBURST99 \citep{leitherer1999}.  
We tune our free parameters $f_{*}$ and $f_{w}$ to simultaneously match the evolving observed Ultra-violet luminosity function (UVLF) and stellar mass function (SMF) at high-$z$ ($z \sim 5-10$). Given that we are not correcting for possible effects such as dust attenuation when tuning our model to observations, we emphasize that our free parameters can be viewed as \textit{observed} ones.
Calibrating our model roughly requires values of $f_{*}=0.03$ and $f_{w}=0.10$.
The resulting UVLFs and SMFs are shown in Fig. \ref{uvluminosityfunction} and \ref{stellarmassfunction}. We note a slight over-prediction in the number density of the rarest brightest $z\lsim 6$ galaxies, which might be a consequence of ignoring the effects of dust, that likely increases with time for massive galaxies. Finally, within our fitting range to the observed data ($\mathrm{M_{UV}}\gsim -23$), the AGN contribution to the UVLF is likely subdominant \citep[e.g][]{ono2018}.

\subsection{The redshift and mass evolution of the central gas density and density profile}
\label{sec_cenden}

We start with the $Ansatz$ that each dark matter halo follows a Navarro-Frenk-White (NFW) density profile \citep{Navarro1996}. This is characterized by a shallow inner ($\propto r^{-1}$) and a steeper ($\propto r^{-3}$) outer density profile; the transition between these regimes is determined by the break radius $r_{s} = r_{\mathrm{vir}} c_h^{-1}$ where $r_{\mathrm{vir}}$ is the virial radius and $c_h$ is the concentration parameter. We use the analytical approximation for $c_h$ as a function of mass and redshift obtained from \citet{prada2012}. In particular, we use Eq. 12-23 from their work to compute $c_h$, and make the assumption that these relations hold below the mass range probed in their work.
Further, assuming baryons to be distributed as an isothermal spherical cloud in hydrostatic equilibrium, the gas density profile can be written as \citep{makino98}   
\begin{equation}
\rho_{g}(x) = \rho_{o}e^{-\gamma^{2}V^{2}}(1+c_h x)^{\frac{\gamma^{2}V^{2}}{x c_h }}.
\label{gasprofile}
\end{equation}
Here, $x = r/r_{\mathrm{vir}}$, $\rho_{o}$ is the central gas density and $\gamma^{2} = 2c_h/F(c_h)$ where $F(c_h) = ln(1+c_h) - c_h/(c_h+1)$. Further, $V^{2}=T_{\mathrm{vir}}/T_{\mathrm{gas}}$ with $T_{\mathrm{vir}}$ and $T_{\mathrm{gas}}$ representing the virial and gas temperatures, respectively. Requiring that the integrated density adds up to the initial gas mass, the central gas density can be written as
\begin{equation}
 \rho_{g}^{o} = \frac{M_{g}^{i} e^{\gamma^{2}V^{2}}   }{ 4\pi  r^{3}_{\mathrm{vir}}} \left[\int^{1}_{0} x^{2}(1+c_h x)^{\frac{\gamma^{2}V^{2}}{xc_h}}dx\right]^{-1}.
\label{cendeneq}
\end{equation}
At any radius, the gas mass density can be translated to a hydrogen number density such that $n_{\mathrm{H}}(x)= 0.92 \rho_{g}(x)[\mu m_{p}]^{-1}$, with $\mu=1.22$ being the mean molecular weight of a neutral primordial gas composed of H and He \citep[][]{barkana2001} and 0.92 represents the fraction of hydrogen atoms\footnote{The central hydrogen gas densities in our low-mass galaxies go down to $n_{\mathrm{H}}\sim 10^{-1.5}\mathrm{cm}^{-3}$, arguably insufficient to form stars. We have explored a scenario in which we have added a dense central hydrogen gas cloud in low-mass galaxies, to find that this is likely not affecting our general trends.
\label{cendenfootnote}}. In the {\it fiducial} model, we further assume that $T_{\mathrm{gas}} = min[T_{\mathrm{vir}}, 2\times 10^{4}\mathrm{K}]$. This results in a situation where low-mass galaxies (with $T_{\mathrm{vir}} < 2\times 10^{4}\mathrm{K}$) have a value of $V^2 =1$ while for high mass galaxies $V^2$ keeps rising since $T_{\mathrm{gas}}$ saturates at $2\times 10^{4}\mathrm{K}$\footnote{Gas can cool to temperatures much lower than our temperature threshold, leading to higher densities. Having explored the addition of dense gas in low-mass galaxies, we follow the same argumentation as the previous footnote, that this would not affect our general trends.
}. 

\begin{figure*}
\center{\includegraphics[scale=0.5]{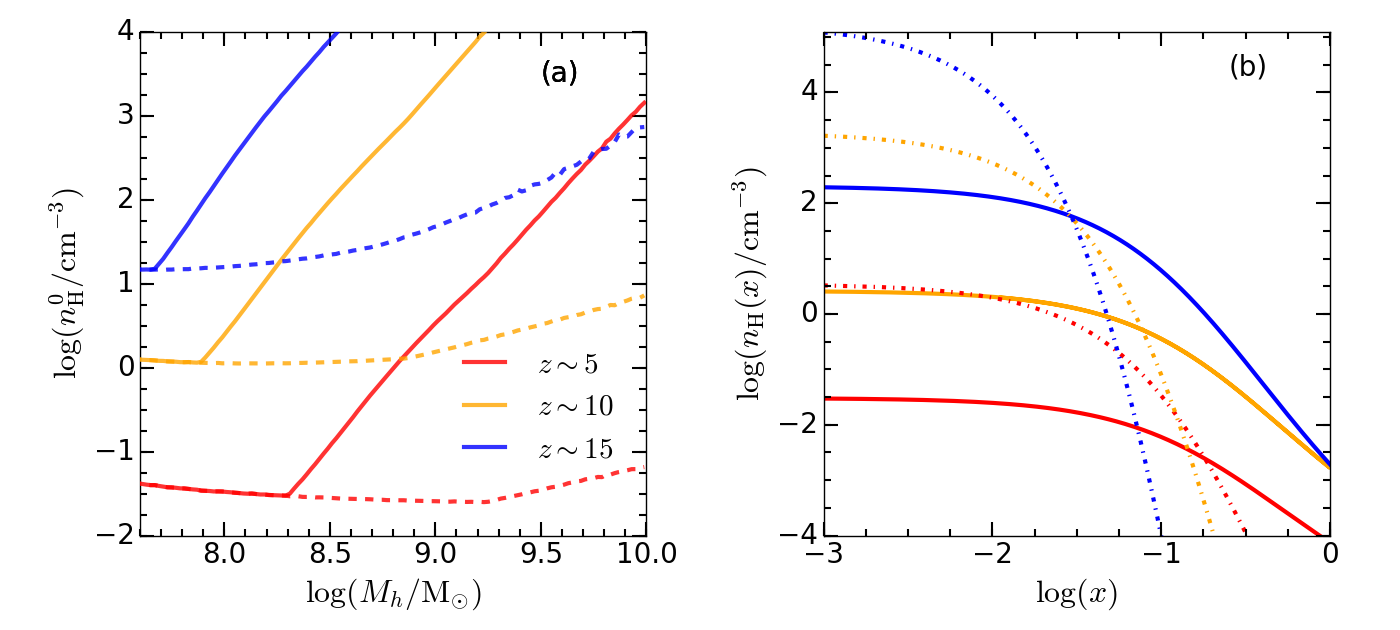}}
\caption{At $z\sim5$ (red), $10$ (yellow) and $15$ (blue), we show results for the average central hydrogen number density as a function of halo mass in the {\it Panel (a)}. The dashed lines indicate the case where $T_{\mathrm{gas}}=T_{\mathrm{vir}}$ and the solid lines highlight the assumption of a temperature threshold at $T_{\mathrm{gas}}= 2\times 10^{4}\mathrm{K}$. The {\it Panel (b)} shows the gas density profile for a $10^{8}$ (solid lines) and $10^{9}\mathrm{M_{\odot}}$ (dotted lines) halo for the case of an assumed temperature threshold at $T_{\mathrm{gas}}= 2\times 10^{4}\mathrm{K}$.} 
\label{cenden}
\end{figure*}

In panel (a) of Fig. \ref{cenden} we show the resulting central hydrogen gas number density $\nho$ as a function of halo mass and redshift as obtained from {\sc Delphi}. As seen, assuming $T_{\mathrm{gas}} = T_{\mathrm{vir}}$ results in a value of $\nho$ that remains roughly constant (increases by a factor of $\sim 3$) over $M_{h}=10^{8-9}\mathrm{M_{\odot}}$ at $z\sim 5 ~(15)$.
However, in the {\it fiducial} model, $\nho$ increases by $\sim 2$ $(3)$ orders of magnitude at $z\sim 5$ $(15)$ which is driven by the increase in the $V^2$ term (Eq. \ref{cendeneq}) with increasing halo mass. Further, due to an increase in the virial temperature with redshift for a given halo mass, this increase shifts to lower masses with increasing redshift. To quantify: a virial temperature of $2\times 10^{4} \mathrm{K}$ corresponds to $M_{h}\simeq10^{8.3}$  ($10^{7.7}$)$\mathrm{M_{\odot}}$ at $z\sim 5$ (15). We also note a substantial corresponding increase in the central density with redshift. For a $10^{9}\mathrm{M_{\odot}}$ halo for example, the central density increases by nearly 5 orders of magnitude from $z\sim 5$ to 15. This increase is primarily driven by the increasing compactness of galaxies: for a given halo mass, the virial radius increases with redshift as $\propto (1+z)^{3}$.

In panel (b) of the same figure, we show the gas density profiles for a $10^{8}$ and $10^{9}\mathrm{M_{\odot}}$ halo at $z\sim5,10$ and $15$ for the saturated temperature case. First of all, we find that the shape of the density profile is very similar at $z \sim 5$ and 10 and not strongly affected by redshift for the $10^8 \msun$ halo; the density decreases by about 3 orders of magnitude between 
$\mathrm{log}(x)=-3$ and $\mathrm{log}(x)=0$. The profile shapes of $10^{9}\mathrm{M_{\odot}}$ halos are significantly more concentrated compared to $10^{8}\mathrm{M_{\odot}}$ halos: indeed, the density falls off by about 3 orders of magnitude within $x=0-0.1$ for such a halo. We also note that the impact of redshift on the profile shape is significantly stronger for a $10^{9}\mathrm{M_{\odot}}$ halo. The density contrast between $x=10^{-3}$ and $x=0.1$ for a $10^{9}\mathrm{M_{\odot}}$ at $z\sim 5$ is roughly 2 orders of magnitude compared to more than 4 orders of magnitude at $z\sim 10$. This is driven by the $V^{2}$ term in Eq. \ref{gasprofile}.

\subsection{Modelling the escape fraction of hydrogen ionizing photons}

We now use the above model to calculate the escape fraction of hydrogen ionizing photons assuming two different scenarios that are ionization-bounded or ionization-bounded with holes as detailed in Secs. \ref{iboundsec} and \ref{denboundsec} that follow. 
\subsubsection{Ionization-bounded scenario}
\label{iboundsec}
We start with our \textit{fiducial} scenario in which LyC leakage occurs when the ionization front (IF) of a galaxy exceeds its virial radius, as proposed in \citet{ferarra2013}. The time-evolution of the IF radius ($r_{\mathrm{I}}$) is set by a detailed balance between the rate at which ionizing photons are produced by the central source and the total recombination rate within the volume enclosed by the IF such that

\begin{equation}
	4\pi r_{\mathrm{I}}^{2}n_{\mathrm{H}}(r_{\mathrm{I}}) \frac{dr_{\mathrm{I}}}{dt} = Q(t) - \int^{r_{\mathrm{I}}}_{0} 4\pi r^{2}n_{\mathrm{H}}(r)^{2}\alpha_{\mathrm{B}}(T_{\mathrm{gas}})dr. 
	\label{balance}
\end{equation}
On the RHS, $Q(t)$ represents the source term (the rate at which ionizing photons are produced by the central source) and the second term represents the recombination rate. Here, $n_{\mathrm{H}}(r)$ is the hydrogen number density at radius $r$ and $\alpha_{\mathrm{B}} (T_{\mathrm{gas}}) = 2.6\times10^{-13}(T/10^{4}\mathrm{K})^{-0.5}$ is the Case B recombination coefficient \citep[][]{ferarra2013}. For the remainder of this work we parametrize $r_{\mathrm{I}}$ in terms of the virial radius as $\Gamma=r_{\mathrm{I}}/r_{\mathrm{vir}}$. When solving Eq. \ref{balance}, we assume a starting radius of $\Gamma=10^{-3}$ for convergence.

We follow the evolution of the IF within each galaxy for 30 Myr, the time spacing of our dark matter merger trees. The time at which the IF breaks out of the virial radius (i.e. $\Gamma=1$) then yields the {\it instantaneous} escape fraction as 
\begin{equation}
	f_{\mathrm{esc}}= \frac{\mathrm{N_{ion}^{esc}}}{\mathrm{N_{ion}^{int}}} =  \frac{ \int_{t(\Gamma=1)}^{30\mathrm{Myr}} Q(t')dt }{ \int_{t=0}^{30\mathrm{Myr}}  Q(t')dt}.
\end{equation}
Here, $\mathrm{ N_{ion}^{esc}}$ and $\mathrm{ N_{ion}^{int}}$ represent the escaping and intrinsic number of hydrogen ionizing photons, respectively.

\subsubsection{Ionization-bounded scenario with holes}
\label{denboundsec}
In this scenario, in addition to the IF, ionizing photons can simultaneously escape through channels cleared of gas due to SNII feedback. We start by calculating the ages of stars (of mass $M_s$) that can explode as SNII as \citep{padovani1993}
\begin{equation}
	t_{\mathrm{SN}} = \left( 1.2 \times 10^{3} \left(\frac{M_{\mathrm{s}}}{\mathrm{M_{\odot}}}\right)^{-1.85} +3   \right) ~\mathrm{Myr}.
	\label{Msn}
\end{equation}
Here, stars with a mass of $M_s = 8$ and $100\msun$ explode as SNII roughly 28.6 and 3.2 Myr after the burst of star formation, respectively. This can be used to infer the SNII rate ($\nu'$) at any time $t'$ within the 30 Myr timestep. Assuming each SNII explosion to release $E_{51}=10^{51}$erg of energy and $f_{w}=0.10$ of this energy coupling to the ISM gas, the energy released by SNII at any time $t'$ within a timestep is 
\begin{equation}
E_{\mathrm{SN}}(t') = f_{w}E_{51} \nu' M_{*}^{\mathrm{new}}(t),
\label{esn}
\end{equation}
where $M_*^{\mathrm{new}}(t)$ is the total new stellar mass formed within a given timestep. Further, the total binding energy at time $t'$ within a given timestep is
\begin{equation}
	E_\mathrm{bin}(t') = M_{g}(t')v_{c}^{2},
	\label{ebin}
\end{equation}
where $v_c$ is the halo circular velocity and $M_g(t')$ is the gas mass at time $t'$ accounting for that lost in ejection. The gas mass ejected at $t'$ can then be calculated as
\begin{equation}
M_{\rm{g}}^{\rm{ej}}(t') = M_g(t') \frac{E_{\mathrm{SN}}(t')}{E_{\mathrm{bin}}(t')} \end{equation}
By the end of the timestep, the total gas mass ejected is
\be
M_{g}^{\rm ej}(t) = [M_g^{\rm i}(t)-M_*(t)] \bigg(\frac{f_*^{\rm eff}}{f_*^{\rm ej}}\bigg).
\ee

\begin{figure*}
\center{\includegraphics[scale=0.5]{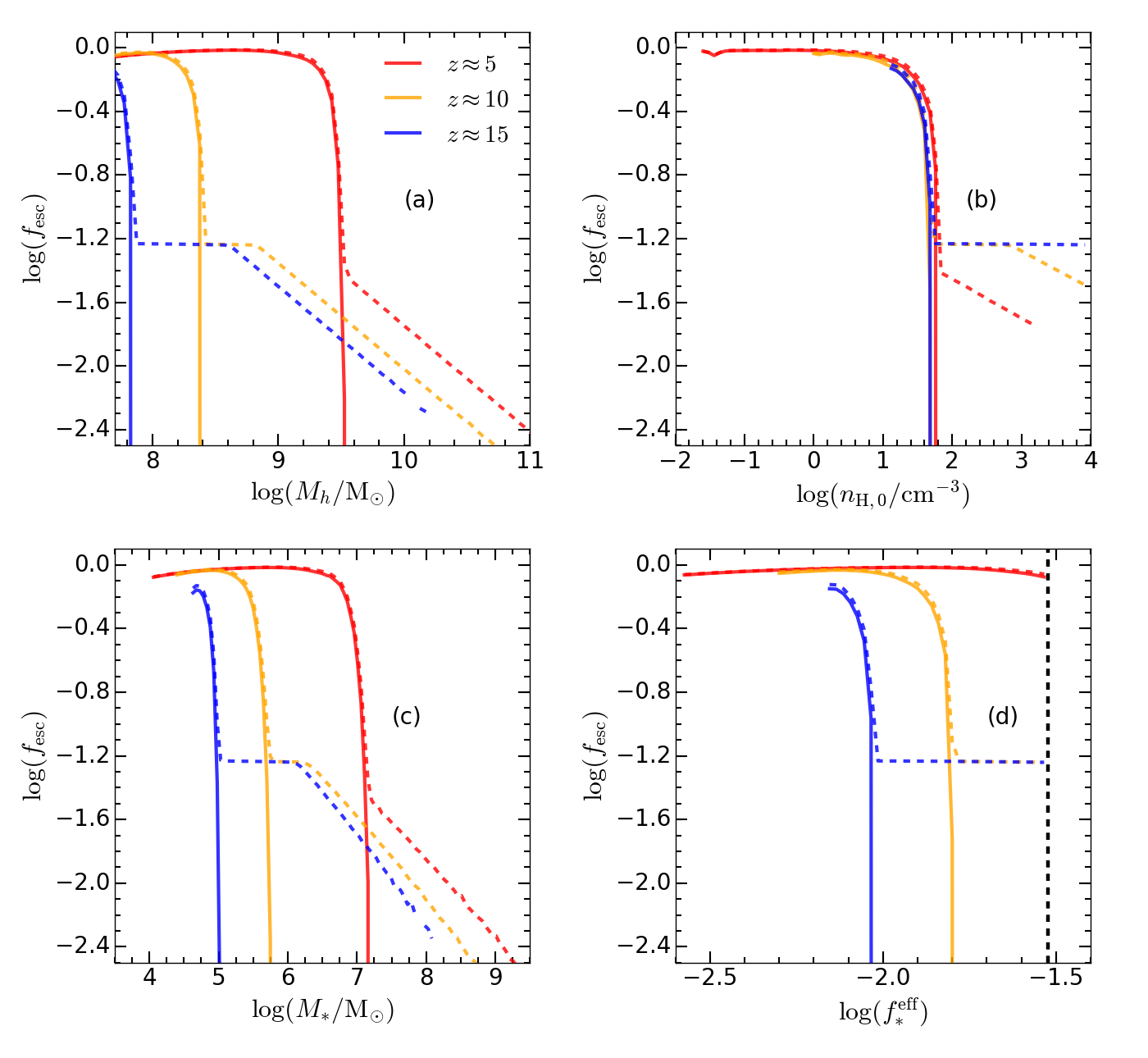}}
\caption{The instantaneous escape fraction binned in terms of halo mass (panel a), central hydrogen gas number density (panel b), stellar mass (panel c) and the star formation efficiency (panel d).
We show our results for $z\sim 5, ~10$ and 15 as highlighted by the different colors, for galaxies with $M_{h}\gsim 10^{7.7}\mathrm{M_{\odot}}$. The solid (dashed) lines indicate the ionization bounded (ionization bounded+holes) scenario. Finally, the vertical (black) dashed line in panel d represents our star formation threshold value of $f_* = 0.03$.} 
\label{fesctrends}
\end{figure*}

Assuming the initial gas mass to be spherically symmetrically distributed within the halo, the fraction of the total spherical solid angle ($4\pi$) that is cleared of gas at time $t'$ is then given by 
\begin{equation}
	\frac{\Omega(t')}{4\pi} = \frac{M_{\mathrm{g}}^{\mathrm{ej}}(t')}{M_{g}^{i}(t)}
	\label{openingangle}
\end{equation} 
Once cleared, we assume the channels to remain open for the rest of the merger tree timestep \footnote{The timescale for which SNII feedback becomes substantial, is significantly longer compared to the timescale at which the majority of ionizing photons are produced. Relaxing this assumption would not change the escape fraction.}. We also assume the gas density profile to be unchanged in the regions not affected by SNII feedback. 

The {\it instantaneous} escape fraction is then given by the combined contribution from both the leaking mechanisms (IF and SNII driven channels) and is computed as
\begin{equation}
	f_{\mathrm{esc}} = \frac{\int^{30\mathrm{Myr}}_{t=0} \frac{\Omega(t')}{4\pi}Q(t') dt + \int_{t(\Gamma=1)} \left[1-\frac{ \Omega(t')}{4\pi}\right] Q(t') dt }{ \int_{t=0}^{30\mathrm{Myr}}  Q(t')dt}.
\end{equation}

\section{The co-evolution of galaxy assembly and the associated escape fraction}
\label{fescsection}
We now study the dependence of the escape fraction on galaxy properties to determine its key drivers in Sec. \ref{trendsection}. We then study the connection between the assembly histories of galaxies and the associated evolving escape fraction in Sec. \ref{fescassem}. Finally, we discuss the time-evolution and variability of the escape fraction in Sec \ref{varsec}.

\subsection{The dependence of $f_{\mathrm{esc}}$ on intrinsic galaxy properties}
\label{trendsection}
In Fig. \ref{fesctrends} we show our results for the instantaneous escape fraction at $z \sim 5$, 10 and 15 for both the models discussed above (ionization bounded and ionization bounded with holes).

We start by discussing the trend with halo mass for the ionization bounded case, shown in panel (a) of Fig. \ref{fesctrends}. At $z\sim 15$, the escape fraction has a maximum value of about $\fesc \sim 0.70$ for $M_h \sim 10^{7.6} \msun$ halos and steeply declines to zero around $M_{h} \simeq 10^{7.8}\mathrm{M_{\odot}}$. This is because galaxies with $M_{h} \gsim 10^{7.8} \msun$ reach a critical central density (corresponding to $\nho\sim 60 \mathrm{cm^{-3}}$) above which the IF can not break out of the virial radius as discussed in panel (b) that follows. At such densities, the recombination rate becomes higher than the ionizing photon production rate. Since $\nho \propto (1+z)^{3}$ (see Fig. \ref{cenden}), the halo mass at which the critical central density is reached corresponds to a larger value of $M_{h} \simeq 10^{8.4}\mathrm{M_{\odot}}$ by $z\sim 10$. This results in galaxies up to this mass being LyC leakers. Further, as a consequence of lower recombination rates at these densities, the IF is sustained beyond $r_{vir}$ for a longer duration, such that values up to $\fesc \simeq 0.93$ are reached for $M_{h} \lsim 10^{8.1} \msun$ halos. The same trends persist at $z\sim5$ where galaxies up to $M_{h} \lsim 10^{9.5} \msun$ are LyC leakers, with $\fesc$ values up to $\sim 0.96$ for $M_{h} \lsim 10^{9.0}\msun$ halos. In the {\it fiducial} model, the value of $\fesc$ decreases with decreasing mass. For example, at $z\sim 5$, $\fesc$ decreases from $\sim 0.95$ at $M_{h} \simeq 10^{8.3}\mathrm{M_{\odot}}$ to 0.83 for $M_{h} \simeq 10^{7.5}\mathrm{M_{\odot}}$ halos. This is the result of a combination of the star formation efficiency that reduces by a factor $\simeq 3.5$ from  $M_{h} \simeq 10^{8.3}\mathrm{M_{\odot}}$ to $\simeq 10^{7.5}\mathrm{M_{\odot}}$ and a central density that remains nearly flat across this mass range at $z\sim 5$. 

Considering the ionization bounded model with holes, we find that the only impact is on halos approaching the critical central density. This is because the IF reaches $r_{\mathrm{vir}}$ on the order of Myrs, before a substantial part of the galaxy has been cleared of gas due to SNII feedback. Hence, by the time the opening angle of SNII-driven gas-free channels starts becoming significant, the production rate of ionizing photons has already dropped substantially. This effect therefore only affects the LyC leakage for galaxies that are too dense to become fully ionized. At $z\sim 15$ we find the escape fraction to remain constant at roughly 0.06 from  $\simeq 10^{7.9}\mathrm{M_{\odot}}$ to $\simeq10^{8.6}\mathrm{M_{\odot}}$. The declining escape fraction with mass for $M_{h} \gsim 10^{8.6}\mathrm{M_{\odot}}$ galaxies can be understood in light of Eq. \ref{openingangle}. As more massive galaxies retain more of their gas mass, the opening angle will be reduced, resulting in a lower escape fraction. At $M_{h}\sim 10^{9.7}\mathrm{M_{\odot}}$, for example, $f_{\mathrm{esc}}$ is reduced to $\sim 0.01$. The same trends persist at lower redshifts: at $z\sim 10$, $\fesc \sim 0.06$ for $M_{h}\simeq 10^{8.4-8.8}\mathrm{M_{\odot}}$. The importance of this model however decreases with redshift: by $z \sim 5$, there is a minimal enhancement in $\fesc$ for $M_h \gsim 10^{9.6}\msun$ halos. This is because galaxies of a given halo mass have deeper potentials with increasing redshift. This leads to a larger fraction of the gas mass being retained resulting in smaller opening angles. 

In panel (b) of the same figure, we show the dependence of $\fesc$ on $\nho$ where we reiterate that $\fesc \sim 0$ at $\nho \gsim 60 \mathrm{cm^{-3}}$ in the {\it fiducial} model. We find this critical value for the central density to remain roughly constant with $z$. We however note that a given value of the central density corresponds to lower halo masses with increasing redshift. Finally, we note that the ionization bounded model with holes leads to an enhancement of $\fesc$ at densities above this critical value; this impact decreases with decreasing redshift as noted above. 

In terms of the $\fesc$ evolution with stellar mass (panel c of the same figure), we again find the same trends as noted in panel (a). This is perhaps not surprising given that our model yields an almost linear relation linking the stellar and halo mass whose normalization increases with increasing redshift. In the {\it fiducial} model, we find LyC leakage to be confined to galaxies with $M_{*}\simeq 10^{5.0}\mathrm{M_{\odot}}$ at $z\sim 15$, while at $z\sim 5$ this mass range extents up to $M_{*}\simeq 10^{7.2}\mathrm{M_{\odot}}$. We also find LyC leakers to extend to lower stellar masses with decreasing redshift. This is because galaxies of a given stellar mass are hosted by progressively massive halos with decreasing redshift. As expected, LyC leakage (albeit with values of $\fesc \lsim 6\%$) extends to galaxies with $M_* \sim 10^6 ~ (10^{6.2}) \msun$ at $z \sim 15 ~ (10)$ considering the ionization bounded model with holes. 

Finally, in panel (d) of Fig. \ref{fesctrends} we show $\fesc$ as a function of the effective star formation efficiency. For both leaking scenarios, we find all galaxies with high escape fractions to be feedback limited ($f_{*}^{\mathrm{eff}}<0.03$) at $z\gsim 7$. With increasing redshift, we find the steep decline in $\fesc$ to occur at lower star formation efficiencies. This is because $f_{*}^{\mathrm{eff}}$ is reached by halos of decreasing mass with increasing redshift. This leads to a weak reduction in the critical central density for leakage at higher redshifts, as visible in panel (b). Interestingly, at a fixed value of $f_{*}^{\mathrm{eff}}$, we find $f_{\mathrm{esc}}$ to decrease with increasing redshift. For example, in the ionization bounded case, at $f_{*}^{\mathrm{eff}} \simeq 7.7\times 10^{-3}$ we find an escape fraction of $\sim 0.66$ $(0.95)$ at $z \sim 15$ $(5)$. This decline is the result of galaxies being denser at higher redshifts for a given star formation efficiency. 

To summarize, we find the combination of the effective star formation efficiency and the central density to set the boundary between galaxies having high and low escape fractions. High escape fractions are found for galaxies that can fully ionize their gas distribution, with gas-free channels created by SNII feedback enhancing the LyC escape from higher mass halos that have a comparatively low star formation efficiency given their central densities.

\begin{figure*}
\center{\includegraphics[scale=0.415]{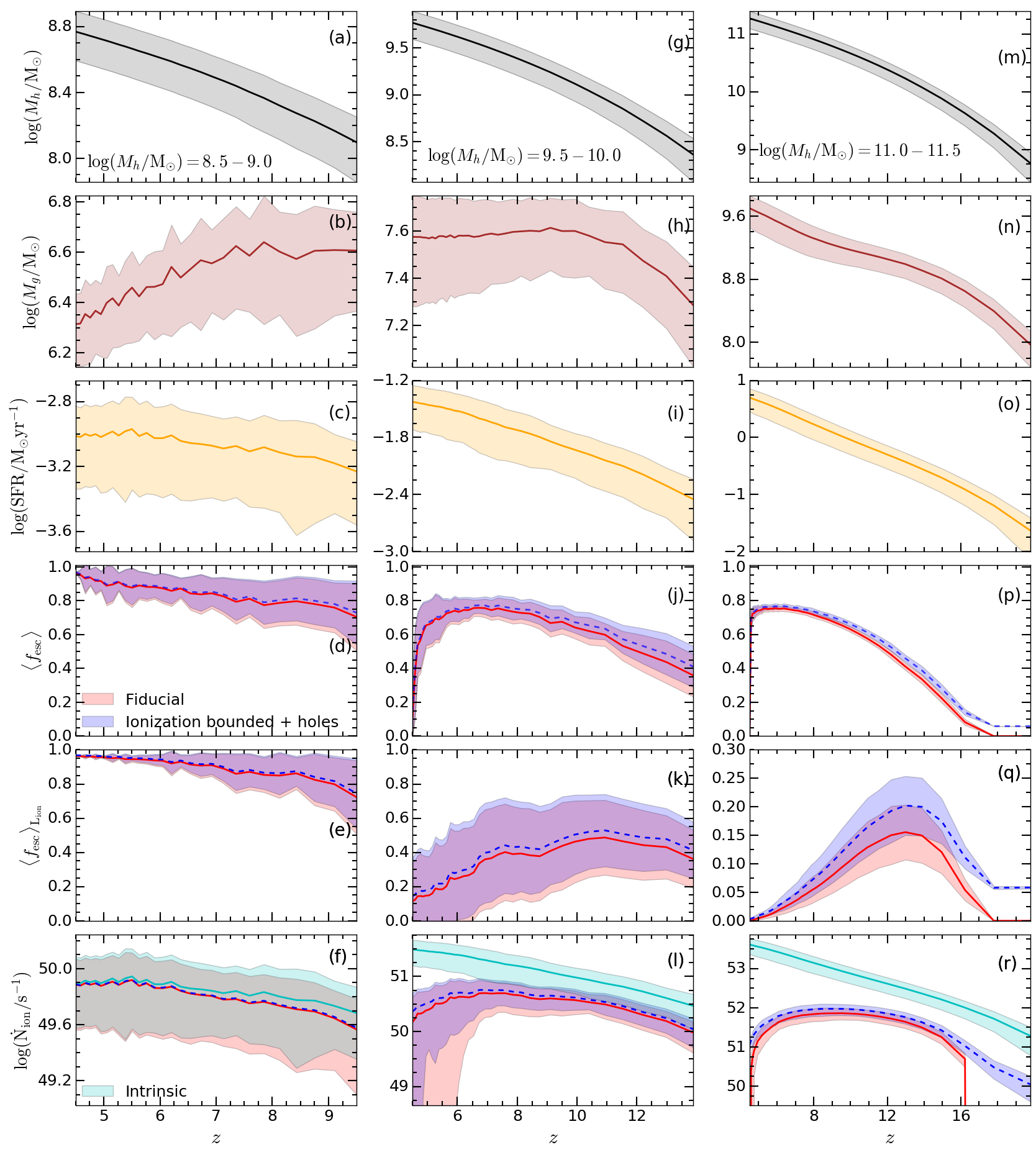}}
\caption{Average assembly histories of low-mass ($M_{h} =10^{8.5-9.0} \msun$ halos; left column), intermediate mass ($M_{h} =10^{9.5-10.0} \msun$ halos; central column) and high-mass ($M_{h} =10^{11-11.5} \msun$ halos; right column). The results in each panel are obtained by averaging over all 50 individual assembly histories in each bin with the shaded region showing the $1-\sigma$ scatter. From top to bottom, we show the redshift evolution of the (a) halo mass, (b) gas mass, (c) star formation rate, (d) average instantaneous escape fraction, (e) fraction of the total intrinsic ionizing photon production rate escaping, and (f) the intrinsic production and escaping rate of ionizing photons. For the last three panels, we show results for both the {\it fiducial} and the ionization bounded with holes models. 
}
\label{avhistories}
\end{figure*}

\subsection{The mass assembly of early galaxies and its associated $f_{\mathrm{esc}}$}
\label{fescassem}
We now study the impact of high-$z$ ($z \gsim 4.5$) galaxy assembly on the escape fraction and its time-evolution. We analyse galaxies in three different halo mass bins at $z \sim 4.5$:  $10^{8.5-9.0}\msun$ (low-mass), $10^{9.5-10}\msun$ (intermediate mass) and $10^{11.0-11.5}$ (high mass); the results are averaged over 50 galaxies in each mass bin. 

We start by discussing results for the low-mass bin as shown in panels (a-f) of Fig. \ref{avhistories}. As shown in panel (a), such galaxies 
assemble roughly $50\%$ of their halo mass by $z \sim 7.1$. The halo assembly shows a scatter of about 0.4 dex around the average as a result of the varied assembly histories of such low-mass objects. In terms of the gas mass (panel b), this remains roughly constant with a value of $M_g \sim 10^{6.6} \msun$ down to $z \sim 7.5$. Below this redshift, the average gas mass declines slightly to $\sim 10^{6.3}\msun$ by $z=4.5$ due to SNII feedback; the only channel through which these galaxies acquire gas is therefore by smooth accretion from the IGM which decreases with decreasing redshift. The gas mass assembly again shows a scatter of about 0.4 dex at almost all redshifts due to 
the assembly histories of such low-mass halos. Despite a decrease in the gas mass at $z \lsim 7.5$, the average star formation rate increases with decreasing redshift (albeit with a scatter of about 0.6 dex) as shown in panel (c) of the same figure. For example, the average star formation rate increases slightly from $\sim 10^{-3.23} \mathrm{M_{\odot}~yr^{-1}}$ at $z\sim 9.5$ to about $ 10^{-3.0}\mathrm{M_{\odot} ~yr^{-1}}$ by $z\sim 4.5$. This is because the increase in the halo mass over these redshifts allows star formation with a higher efficiency which dominates over the decrease in the gas mass available for star formation. The average {\it instantaneous} escape fraction shows an increase from about $\avfesc \sim 0.70$ at $z \sim 9.5$ to $\avfesc \sim 0.96$ by $z \sim 4.5$ as an increasingly large number of low-mass (LyC leaking) progenitors form with decreasing redshift, as shown in panel (d) of the same figure. Given that most of the gas mass of such low-mass galaxies is ejected by SNII in our model, the results of both the $\fesc$ models ( {\it fiducial} and ionization bounded model with holes) are extremely similar over the entire assembly.

\begin{figure*}
\center{\includegraphics[scale=0.37]{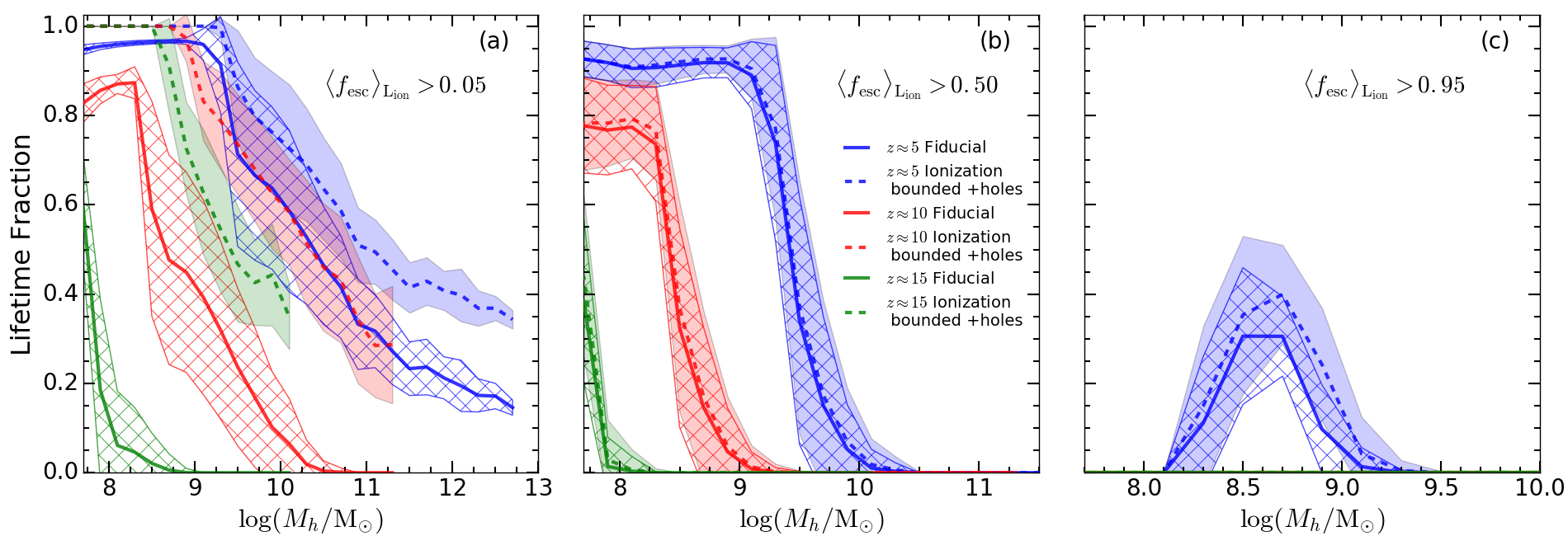}}
\caption{Fraction of the lifetime for which the escape fraction is above the threshold noted as a function of halo mass. For both the leaking models studied in this work (ionization bounded and ionization bounded with holes), the lines show average results for galaxies at $z \sim 5-15$ (as marked) for $\effesc>0.05$ (panel a), $>0.50$ (panel b), and $>0.95$ (panel c). In each panel, the shaded regions show the $1-\sigma$ scatter.} 
\label{lifetimezimpact}
\end{figure*}

We then define the ``ionizing luminosity-weighted" escape fraction over all ($N$) progenitors of a galaxy at redshift $z$ as 
\begin{equation}
    \effesc(z)  = \frac{ \displaystyle \sum_{i=1}^N f_{\mathrm{esc}}(i,z)\dot{\mathrm{N}}_{\mathrm{ion}}^{\mathrm{int}}(i,z)  }{ \displaystyle \sum_{i=1}^N \dot{\mathrm{N}}_{\mathrm{ion}}^{\mathrm{int}}(i,z) }.
\label{lumweighted}    
\end{equation}
This quantity represents the fractional rate of ionizing photons that escape the ISM at redshift $z$. The value of this effective escape fraction increases from $\sim 0.72$ at $z\sim 9.5$ to $\sim 0.96 $ at $z=4.5$ (panel e), for both the {\it fiducial} and ionization bounded model with holes $\fesc$ models, driven by an increase in the {\it instantaneous} $\fesc$ values shown above. Finally, we study the relation between the intrinsic and escaping ionizing photon production rates as shown in panel (f) of the same figure. Given that the number of short-lived massive stars are mainly responsible for the production of ionizing photons, the intrinsic ionizing photon production rate effectively tracks the star formation rate at all $z$. As for the escaping production rate, 
this closely starts tracking the intrinsic rate with decreasing redshift due to the high escape fractions for galaxies in this mass bin. 

Quantitatively, the halo assembly is very similar for the intermediate-mass bins (panel g) of the same figure, with 50\% of the mass assembling by $z \sim 7.4$. However, given the higher potentials associated with such halos, they can keep a significant gas mass bound within them. This leads to a flattening in the gas mass at $z \lsim 10$ at $M_g \sim 10^{7.6}\msun$ as shown in panel h. This also results in the star formation rate increasing with decreasing redshift, from $\sim 10^{-2.5} \mathrm{M_{\odot} ~yr^{-1}}$ at $z \sim 14$ to $\sim 10^{-1.4} \mathrm{M_{\odot} ~yr^{-1}}$ by $z \sim 4.5$ as shown in panel (i). Interestingly, the redshift evolution of the average instantaneous escape fraction is quite different as shown in panel (j) - due to more massive progenitors (that have smaller $\fesc$ values), $\avfesc \sim 0.36$ at $z \sim 14$ which increases to $\sim 0.75$ by $z \sim 7$ as an increasingly larger number of low-mass progenitors form with high $\fesc$ values. However, this quantity shows a drop at $z \lsim 7$ as increasingly massive systems with lower escape fractions assemble. On the other hand, the effective escape fraction is fairly flat with a value of $\effesc \sim 0.4$ at $z \gsim 8$ after which it drops to $\sim 0.1$ by $z \sim 4.5$ as shown in panel (k). This decline at $z\lsim 7$ is again caused by the assembly of high-mass systems with very low escape fraction values. This trend is also evident when comparing the evolving escaping and intrinsic production rate of ionizing photons in panel (l). While the latter increases with decreasing redshift tracing the star formation rate, the former flattens with decreasing redshift and declines at $z\lsim 7$.

Halos in the high-mass bin start assembling much earlier as expected and assemble 50\% of their mass by $z \sim 7.4$ as shown in panel (m). Given their more massive progenitors, for such galaxies the gas mass keeps rising with decreasing redshift, from $M_g \sim 10^{8.0}\msun$ at $z \sim 20$ to $10^{9.7}\msun$ by $z \sim 4.5$ as shown in panel (n). Given their larger number of progenitors, the scatter in the assembly history is naturally lower than that seen in the low and intermediate-mass bins. As a result of the quasi-monotonic growth of gas mass, the star formation rates too grow with decreasing redshift, from $\sim 10^{-1.6} \mathrm{M_{\odot} ~yr^{-1}}$ at $z \sim 20$ to $\sim 10^{0.7} \mathrm{M_{\odot} ~yr^{-1}}$ by $z \sim 4.5$ as shown in panel (o). Their more massive progenitors (as compared to low-intermediate mass halos) also result in lower values of $\avfesc$ - this increases from $\sim 8\%$ at $z \sim 16$ to $\sim 0.75$ by $z \sim 6$ after which it shows a drop as shown in panel (p). The cause of this drop is analogous to that seen in the intermediate mass bin. Dominated by high-mass progenitors with low $\fesc$ values, $\effesc$ too decreases with redshift from $\sim 0.16$ at $z \sim 13$ to roughly $10^{-3}$ by $z \sim 4.7$. While the values are similar for both $\fesc$ models at $z \lsim 15$ (panel q), at higher redshifts, $\effesc \sim 0.06$ in the ionization bounded model with holes (as compared to no leakage in the {\it fiducial} model). This is a consequence of the fact that leakage here is dominated by low mass galaxies, that always have an instantaneous escape fraction of $\sim 0.06$ (Sec. \ref{trendsection}). The growing difference between the intrinsically produced and the escaping number of ionizing photons (panel r) below $z\lsim 13$ is also driven by the assembly of an increasingly large number of high-mass systems with negligible escape fractions.

\subsection{The fractional lifetime spent in high and low-$f_{\mathrm{esc}}$ regimes }
\label{varsec}
\begin{figure*}
\center{\includegraphics[scale=0.35]{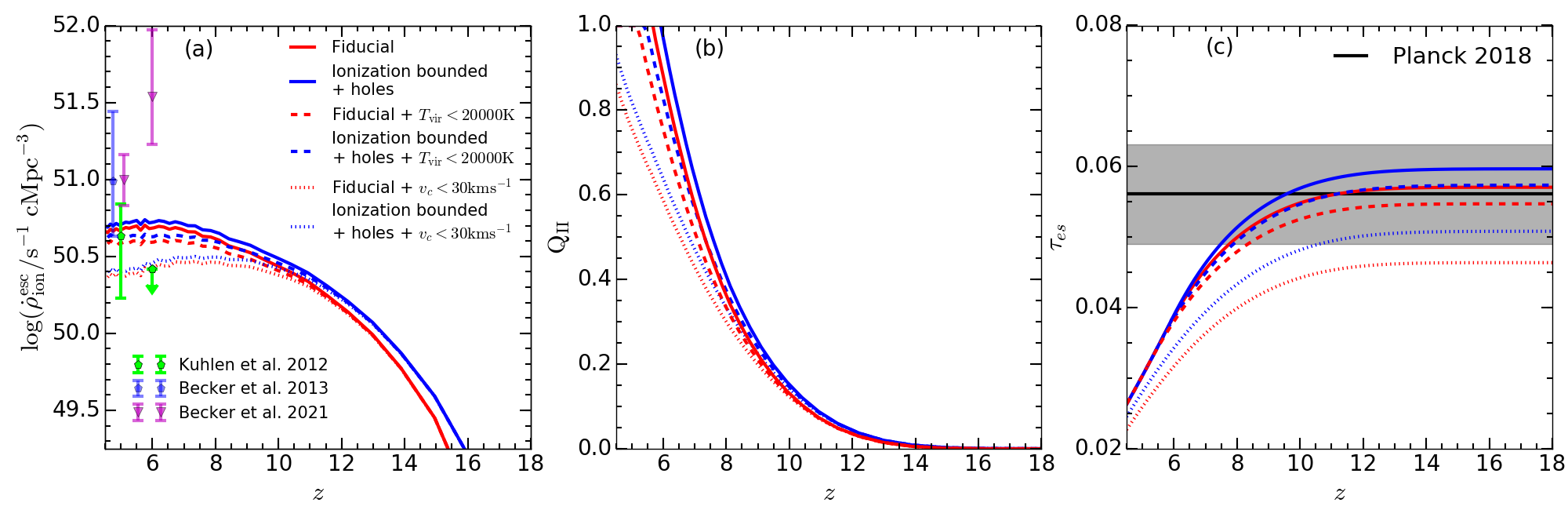}}
\caption{{\it Panel (a)}: Ionizing emissivity as a function of redshift, with observed estimates from \citet[][]{kuhlen2012}, \citet[][]{becker2013} and \citet[][]{becker2021}, as marked. {\it panel (b)}: Volume filling factor of ionized hydrogen, as a function of redshift. {\it panel (c)}: results for the Thomson scattering optical depth, along with the result from \citet[][]{planck2018} and its grey $1-\sigma$ uncertainty band. In each panel, the red and blue \textbf{colors} show the results of the ionization bounded and the ionization bounded model with holes, respectively. The different linestyles indicate the different feedback scenarios; solid lines, no reionization feedback; dashed lines, $T_{\mathrm{vir}}<20000\mathrm{K}$ feedback; and dotted lines, $v_{c}<30\mathrm{kms^{-1}}$ feedback.}

\label{emissivity}
\end{figure*}

We now discuss the fraction of their lifetimes that galaxies spend in different $\effesc$ regimes for both the models studied in this work. Starting with the {\it fiducial} model, as shown in panel (a) of Fig. \ref{lifetimezimpact}, $z \sim 5$ galaxies with $M_{h}=10^{7.7-9}\msun$ show $\effesc \gsim 0.05$ values for almost the entirety ($\gsim 95\%$) of their lifetime. The fractional lifetime for which $z \sim 5$ galaxies show $\effesc \gsim 0.05$ decreases with increasing halo mass, falling to about $46\% ~ (17\%)$ for galaxies with $M_h \sim 10^{10.5} ~ (10^{12.5})\msun$. This trend is driven by the assembly of an increasing number of massive systems with $\fesc=0$, as detailed above. At each redshift, the scatter shows the variety of the assembly histories of these halos that crucially impact their $\effesc$ values. At $z\sim 5,10$ and 15, the mass at which the lifetime spent in the $\effesc \gsim 0.05$ regime drops shifts to $\sim 10^{9.5}, 10^{8.3}$ and $10^{7.8}\mathrm{M_{\odot}}$. This mass is directly related to the critical mass for ionization bounded leakage as discussed in Sec. \ref{trendsection}; as structures start to assemble above this limit, $\effesc$ naturally gets lowered. This in turn leads to a shift in the mass range of maximum variability, since the mass range close to the critical mass for leakage is most sensitive to the underlying mass assembly history. The region in which we encounter the highest variability shifts to $ M_{h} \sim 10^{9.5-10.0}, 10^{8.5-9.0}$ and $10^{7.9-8.3}\mathrm{M_{\odot}}$ at $z\sim 5,10$ and 15. We also find that the fractional lifetime decreases with increasing redshift for a given halo mass. On average, a $10^{9.5}\mathrm{M_{\odot}}$ halo at $z\sim 5$ spends roughly $71\%$ of its lifetime at $\effesc\gsim0.05$ which reduces to $24\%$ at $z\sim 10$; this threshold is never reached at $z\sim 15$. This trend is driven by an increasing fraction of assembled $f_{\mathrm{esc}}=0$ structures at higher redshifts, a result of the declining critical mass for leakage with increasing redshift.

The ionization bounded model with holes mainly affects the lifetimes spent above the low $\effesc \gsim0.05$ threshold as shown in panel (a) of Fig. \ref{lifetimezimpact}; the impact is the strongest at the highest redshifts of $z \sim 15$ and (within the $1-\sigma$ scatter) converges towards the {\it fiducial} model by $z \sim 5$. As discussed in Sec. \ref{trendsection}, feedback limited galaxies always have an escape fraction of $\sim 0.06$. We also find the lifetimes spent in this regime are increasingly affected at larger masses, in line with Sec. \ref{fescassem}. The impact can be clearly seen at $M_{h} >10^{12.5} \mathrm{M_{\odot}}$. 

Moving onto the regime with $\effesc \gsim 0.5$, low-mass galaxies ($M_{h}=10^{7.7-9}\mathrm{M_{\odot}}$) again spend $\gsim 90 \% $ of their lifetime in this regime. However, in the $10^{9.5-9.75}\msun$ mass range, we find the fractional lifetime spent in this regime to show an extremely large range between 0 and 97 $\%$. Finally, more massive systems with $M_h \gsim 10^{10}\msun$ always show $\effesc$ values lower than 50\% due to their high central densities and comparatively low star formation rates. The same trends persist here as shown for the low $\effesc$ regime: at a given halo mass, the fractional lifetime spent in a given $\effesc$ regime decreases with increasing redshift. A clear difference however, as expected, is the rapid decline in the lifetime fraction spent in this regime compared to $\effesc \gsim 0.05$. Finally, there are no significant differences between the results from the {\it fiducial} and ionization bounded with holes models for this regime since galaxies with $\effesc \gsim 0.50$ are predominantly driven by leakage through ionized gas rather than SNII-created channels.

Finally, we find only low-mass galaxies ($M_h \sim 10^{8.2-9.4}\msun$) can emit in the high $\fesc$ regime (with $\effesc \gsim 0.95$). However, these low-mass halos, where the star formation rate allows the IF to break out of the virial radius, spend only about $30\%$ of their lifetime in this high $\effesc$ regime at $z \sim 5$; as a result of their deeper potentials, galaxies do not emit in this regime at the higher redshifts considered here. This lifetime fraction shows a slight average increase of up to about 10\% (although this is within the scatter from the {\it fiducial} model) when the ionization bounded model with holes is considered.  

In the remainder of this work, we focus on the implications of these results in the context of reionization.

\section{The contribution of early galaxies to reionization}
\label{riosection}
We now focus on the implications of our $f_{\mathrm{esc}}$ modelling approach on the epoch of reionization. In Sec. \ref{eorhist} we calculate the reionization history from our model and validate it against global observables. This is then followed by a discussion of the key sources and their contribution to the ionizing background in Sec. \ref{keysources}. 

\subsection{The emissivity and optical depth}
\label{eorhist}
We use the approach discussed in \citet{dayal2017} to calculate the reionization history which can be expressed in terms of the redshift evolution of the volume filling fraction of ionized hydrogen ($Q_{\mathrm{II}}$) such that \citep[e.g.][]{madau1999}
\begin{equation}
\frac{dQ_{\mathrm{II}}}{dz} = \frac{\dot{\rho}^{\mathrm{esc}}_{\mathrm{ion}}}{n_{\mathrm{H}}} \frac{dt}{dz}  -\frac{Q_{\mathrm{II}}}{t_{\mathrm{rec}}} \frac{dt}{dz},
\end{equation}
where the first term on the right hand side accounts for the growth of ionized regions due to the input of ionizing photons and the second term expresses the counteracting effect due to recombinations. Here, $\dot{\rho}^{\mathrm{esc}}_{\mathrm{ion}}$ is the ionizing emissivity (the rate at which ionizing photons are injected into the IGM per unit volume), $n_{\mathrm{H}}$ is the comoving hydrogen density  and $dt/dz = [H(z)(1+z)]^{-1}$. The ionizing emissivity at any redshift is calculated by weighting the emergent ionizing photon rate by the number density of any given galaxy at any $z$. Further, the recombination time can be expressed as \citep[][]{madau1999}
\begin{equation}
t_{\mathrm{rec}} = \frac{1}{ n_{\mathrm{H}} \chi(1+z)^{3} \alpha_{\mathrm{B}}C} ,
\end{equation}
where $\chi = 1.09$ factoring in the extra electrons originating from singly ionized helium and $C =  1+43 ~z^{-1.71}$ is the clumping factor of the IGM \citep{pawlik2009, haardt2012}. The integrated Thomson electron scattering optical depth can then be calculated as  
\begin{equation}
    \tau_{\mathrm{es}}(z) = \sigma_{\mathrm{T}}c \int_{0}^{z} \frac{  n_{e}(z^{'}) (1+z^{'})^{2}}{H(z^{'})} dz^{'},
\end{equation}
where $c$ is the speed of light, $\sigma_{\mathrm{T}}=6.65\times 10^{-25}\mathrm{ cm^{-2}}$ is the Thomson scattering cross section and the comoving averaged global electron number density is computed as $n_{e}(z) = n_{\mathrm{H}}\chi Q_{\mathrm{II}}(z)$, where we assume He to be fully ionized at $z<3$ and singly ionized otherwise \citep[e.g.][]{kulkarni2019}.

We start by comparing the emissivity calculated from our model with observations in panel (a) of Fig. \ref{emissivity}. It is really heartening to see that our emissivity estimates for both $\fesc$ models are in excellent agreement with the observations. Interestingly, our results naturally show a downturn of the emissivity at $z \lsim 6$. This is driven by the downturn in $\fesc$ detailed in Sec. \ref{fescassem}. 

The similarity in the emissivity for both $\fesc$ models results in very similar reionization histories as shown in panel (b) of the same figure. With the appearance of the first LyC leaking galaxies, the epoch of reionization in our {\it fiducial} model starts at $z\sim 16.2$ roughly 240Myr after the big bang. Within the following $\sim 200$Myr down to $z\sim 10.4$, reionization proceeds rather slowly being $10\%$ complete. By a redshift of $\sim 7.4$ or about $700$Myr after the big bang, the universe is roughly $50\%$ ionized. We find reionization to be complete by $z=5.67$; in the ionization bounded model with holes, reionization follows almost the same history, ending roughly 50Myr earlier at $z\sim 5.91$. 

Finally, we note that both $\fesc$ models are compatible with the latest electron scattering optical depth measured by \citet{planck2018} as shown in panel (c) of the same figure. While the results from the {\it fiducial} model lie very close to the central measured value of $\tau_{es} \sim 0.055$, the ionization bounded model with holes yields a slightly higher value of about 0.06. This result is a particularly good sanity check of our model \citep[see also][]{ferarra2013} since once the star formation parameters are fit to galaxy observables (the UV LF and SMF), we have no further free parameters when calculating the $\fesc$ values for any galaxy.

We now explore the effects of reionization feedback on these results.

\subsection{\textbf{The impact of reionization feedback}}
\label{reionizationfeedback}

As re-ionization proceeds, the ultra-violet background (UVB) in the ionized IGM heats to the IGM to temperatures of roughly $1-4 \times10^4$K. Galaxies exposed to this UVB can experience lower gas accretion rates onto dark matter halos \citep{couchman-rees1986, Hoeft2006}, or the removal of their gas content into the IGM through photo-evaporation \citep{barkana-loeb1999, shapiro2004}. Low-mass halos residing in ionized regions are predominantly affected and can lead to a reduced SFR. This however depends on the spatial variation of re-ionization progress \citep[e.g.][]{Hasegawa2013, Gnedin2014, Pawlik2015, Ocvirk2018, Katz2019, Wu2019, Hutter2021a}.
The combined escaping emissivity as a function of redshift arising both from ionized as well as neutral regions can be expressed as \citep[e.g.][]{dayal2017, choudhury2018},  
\begin{equation}
\label{weighemeq}
   \dot{\rho}^{\mathrm{esc}}_{\mathrm{ion, fb}}(z) =  [1- Q_{\mathrm{II}}(z)]  \dot{\rho}^{\mathrm{esc}}_{\mathrm{ion},\mathrm{I}}(z) + Q_{\mathrm{II}}(z)\dot{\rho}^{\mathrm{esc}}_{\mathrm{ion},\mathrm{II}}(z).
\end{equation}

The contribution arising from galaxies residing in neutral regions not affected by feedback is expressed by the first term, while the second term accounts for the contribution of galaxies affected by feedback in ionized regions. In the very early stages of reinization when the IGM is mosly neutral or $Q_{\mathrm{II}}\simeq 0$, such that $\dot{\rho}^{\mathrm{esc}}_{\mathrm{ion, fb}} \sim \dot{\rho}^{\mathrm{esc}}_{\mathrm{ion},\mathrm{I}}(z)$, while towards the completion of reionization 
$Q_{\mathrm{II}}\simeq 1$, leading to an emissivity of $\dot{\rho}^{\mathrm{esc}}_{\mathrm{ion, fb}} \sim \dot{\rho}^{\mathrm{esc}}_{\mathrm{ion},\mathrm{II}}(z)$.
We consider two different reionization feedback scenarios:
\newline
\newline
\indent (i) Suppression of the gas content of galaxies with $T_{\mathrm{vir}}<20000\mathrm{K}$, corresponding to $M_h \sim 10^{8.3}$ ($10^{7.7}\msun$) at $z\sim 5$ ($15$).   
\newline
\indent (ii) Suppression of the gas content of galaxies with $v_{c}<30\mathrm{kms^{-1}}$, corresponding to $M_h \sim 10^{9.1}~(10^{8.45}\mathrm{M_{\odot}}) $ at $z \sim 5~(15)$.   
\newline
\newline
We stress however that in reality the extent of this reionization feedback is likely dependent on the environment in which galaxies reside. Given the absence of spatial information in our semi-analytical model, we are unable to account for this as such.
\newline
\indent In panel (a) of Fig. \ref{emissivity} we show the resulting emissivites for our two feedback scenarios. Att the highest redshifts ($z\gsim 13$), the non-feedback and models including feedback are roughly identical, as expected given the early stage of reionization. As reionization proceeds, the effect of feedback between the different models becomes apparent, as the emissivity starts to diverge from the non-feedback case. For the fiducial leaking model at $z\sim 6$ the emissivity in the case of $T_{\mathrm{vir}}<20000\mathrm{K}$ feedback is roughly $81\%$ of the non-feedback case, while in the stronger $v_{c}<30\mathrm{kms^{-1}}$ feedback scenario, the emissivity is reduced to $\sim 55\%$. This relative decline in emissivity is roughly identical in the ionization bounded + holes leaking model. While both leaking models in the absence of reionization feedback models are compatible with multiple observed datasets, in the case of both feedback models, the emissivity is only compatible with the observed estimates from \citet[][]{kuhlen2012}. 
\newline
\indent The resulting reionization histories are shown in panel (b) of Fig. \ref{emissivity}. With only a small fraction of the universe ionized, the reionization histories at $z\gsim 10$ are nearly indistinguishable between the different reionization feedback cases. On the other hand, the decline in the emissivity due to the suppression of galaxies as reionization proceeds, naturally results in a delay in the reionization process. Comparing at $z\sim 5.9$, in the absence of reionization feedback $Q_{\mathrm{II}}\simeq 0.92$, with $T_{\mathrm{vir}}<20000\mathrm{K}$, $Q_{\mathrm{II}}\simeq 0.78$, and in the $v_{c}<30\mathrm{kms^{-1}}$ case, $Q_{\mathrm{II}}\simeq 0.60$. This delay between the different feedback models is qualitatively the same in the ionization bounded model + holes. We note however that the $v_{c}<30\mathrm{kms^{-1}}$ feedback model fails to reionize the universe \footnote{This feedback model is compatible with the observed UVLFs.}. 
\newline
\indent Lastly, the resulting optical depth for both feedback realizations are shown in panel (c) of Fig. \ref{emissivity}. We find that for both leaking models in the $T_{\mathrm{vir}}<20000\mathrm{K}$ feedback scenario, the resulting electron scattering optical depth is in excellent agreement with the observed value of \citet{planck2018}. While in the fiducial leaking model, the $v_{c}<30\mathrm{kms^{-1}}$ feedback scenario under-predicts $\tau_{\mathrm{es}}$, when leakage through holes is included, the $v_{c}<30\mathrm{kms^{-1}}$ feedback scenario is in accord with the observed $\tau_{\mathrm{es}}$.

\subsection{The key sources of reionization}
\label{keysources}
\begin{figure*}
\center{\includegraphics[scale=0.46]{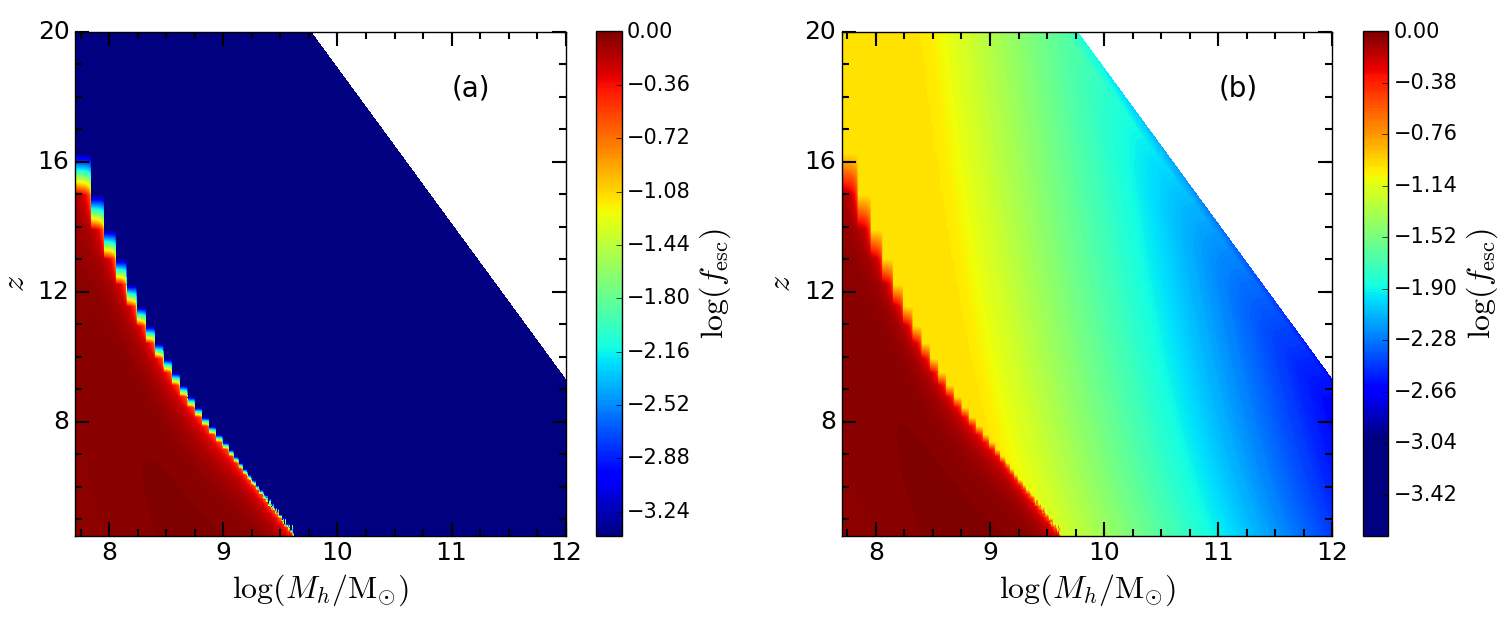}}
\caption{\small 2d map showing the instantaneous escape fraction as a function of redshift and halo mass for the ionization bounded model in panel (a) and the ionization bounded model + holes in panel (b). The values for the escape fraction are represented by the color bar on the right hand side of each panel. The dark blue region in the left panel represents $f_{esc}=0$. For display purposes, we have set the escape fraction to $10^{-3.4}$ when $f_{esc}=0$.} 
\label{2dmap}
\end{figure*}

Based on our calculations, we now discuss the star-forming galaxies that would provide most of the photons for reionization. This naturally depends on three key parameters: (i) the intrinsic production rate of ionizing photons; (ii) the escape fraction from any galaxy; (iii) the number density of the galaxy.

\begin{figure*}
\center{\includegraphics[scale=0.42]{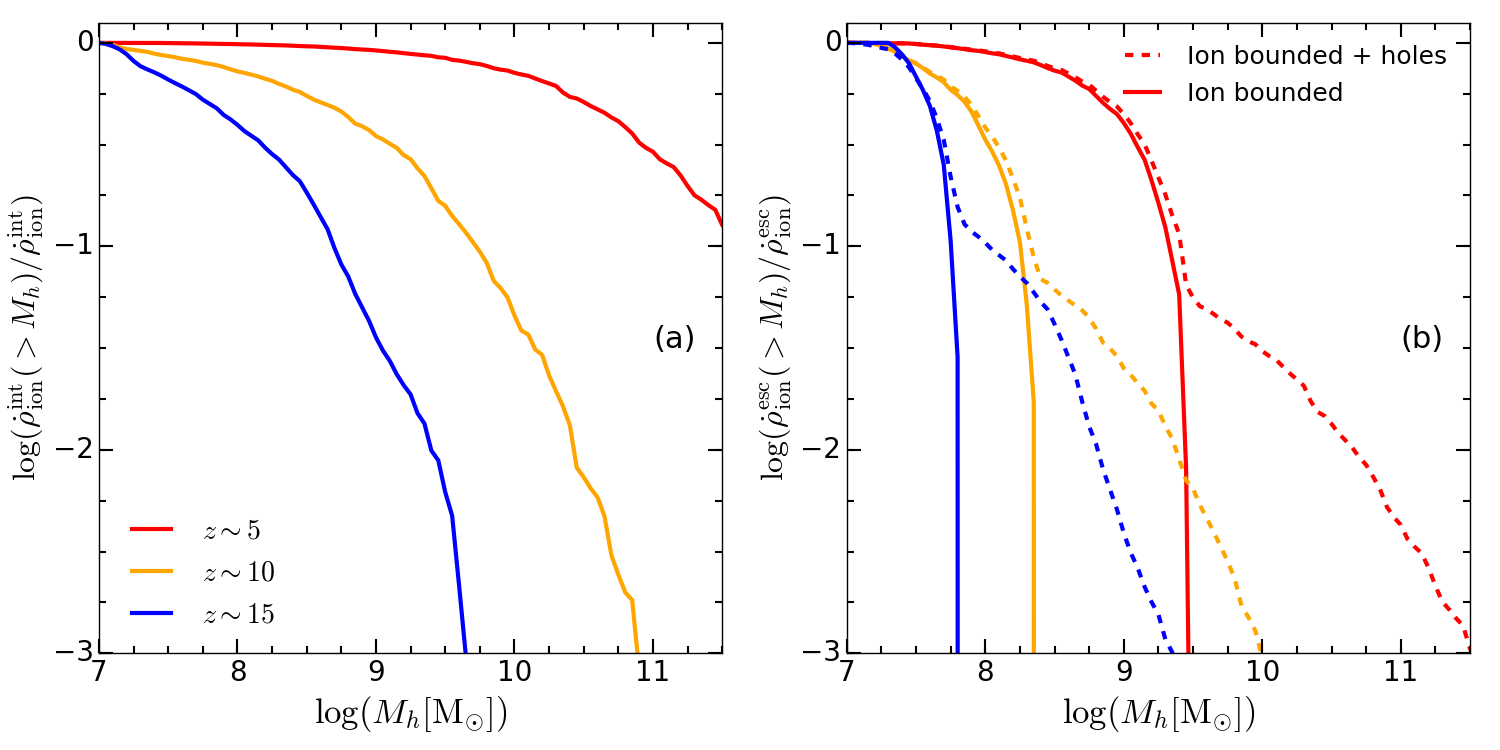}}
\caption{\small  Relative contribution to the ionizing emissivity of galaxies above a given halo mass as a function of halo mass at $z\sim 5$, 10 and 15. In panel (a), we show the results for the intrinsic emissivity, while the results for the escaping emissivity are shown in panel (b). Solid (dashed) lines in the right panel show results for the ionization bounded (+holes) model.} 
\label{eorcontribution}
\end{figure*}

To this end, we start by showing the dependence of the instantaneous escape fraction on mass and redshift for all galaxies in 
the ionization and density bounded models in Fig. \ref{2dmap}. For the {\it fiducial} ionization bounded model shown in panel (a), LyC leakers are limited to low-mass galaxies ($M_h \lsim 10^{7.8}\msun$) at $z \sim 15$ that show $\fesc$ up to $\sim 0.7$. Due to the decrease in the central gas density with decreasing redshift as discussed in Sec. \ref{sec_cenden}, galaxies with masses as large as $M_h \sim 10^{9.5} \msun$ show similar $\fesc$ values by $z \sim 5$. As has been discussed in the previous sections, higher mass galaxies effectively show no leakage of ionizing photons. Considering the ionization bounded model with holes naturally increases the contribution from higher mass halos at all redshifts as shown in panel (b) of the same figure. The $\fesc$ values are effectively unaffected for the LyC leakers in the {\it fiducial} model, as might be expected, given that the escape in this case is driven by the IF breaking out of the virial radius. However, given the additional contribution to escape from SNII-channels, in this case, galaxies as massive as $M_h \sim 10^{9.8}\msun$ show $\fesc \gsim 1\%$ at $z \sim 20$. By $z \sim 5$, galaxies of $M_h \sim 10^{10-12} \msun$ show $\fesc$ values ranging between $0.1-2\%$. 

We then compute the fraction of the emissivity arising from galaxies above a given mass with respect to the total emissivity as shown in Fig. \ref{eorcontribution}, for the case of no reionization feedback. In terms of the intrinsic emissivity (panel a of this figure), at $z \sim 15$, roughly 60\% of the ionizing photon density is provided by low-mass galaxies with $M_h \lsim 10^8 \msun$. Although the star formation rates increase with increasing mass, the drop in the number density results in galaxies with $M_h \gsim 10^9 \msun$ providing less than 5\% to the intrinsic emissivity. As increasingly massive systems assemble with decreasing redshift, 50\% of the intrinsic emissivity comes from systems as massive as $M_h \gsim 10^{8.6} ~ (10^{10.5})\msun$ by $z \sim 10 ~ (5)$.

Starting with the ionization bounded model, at $z\sim 15$ we find that, as a result of their high $\fesc$ values, roughly half of the ionizing emissivity is contributed by galaxies with $M_{h}\gsim 10^{7.6}\msun$. As might be expected, the galaxies providing 50\% of the ionizing emissivity shift to progressively higher halo masses with decreasing redshift. Indeed, half of the escaping ionizing emissivity comes from halos with $M_h \gsim 10^{7.9} ~ (10^{8.9})\msun$ by $z \sim 10 ~ (5)$.  

We now compare these results to the ionization bounded model with holes. As might be expected, LyC leakers (where the IF can break out of the virial radius) are not very much affected by the additional escape from gas-free channels. Indeed, as seen, the galaxies that provide 50\% of the escaping ionizing emissivity are hardly affected by the presence of SNII-cleared channels. However, these channels affect the tail end of leakers (that can provide up to 13\% of escaping photons) significantly. Galaxies that are more massive compared to the {\it fiducial} model can provide the last few percent of reionization photons when SNII channels are included.
Quantitatively, leakage through SNII channels from $M_{h}\gsim 10^{9.5}\mathrm{M_{\odot}}$ galaxies contribute $6\%$ to reionization at $z \sim 5$, which increases to $13\%$ by $z \sim 15$ for $M_{h}\gsim 10^{7.8}\mathrm{M_{\odot}}$ galaxies. Lastly, we note that in the case of reionization feedback, as an increasingly large population of low-mass galaxies is suppressed, the fractional contribution to the escaping ionizing emissivity of more massive galaxy naturally increases.

Finally, we then define a ``cumulative effective escape fraction" for each galaxy as 
\begin{equation}
    f^{\mathrm{cum}}_{\mathrm{esc}}(z) = \frac{ \mathrm{N_{ion, cum}^{esc}}(i,z')   }{\mathrm{N_{ion,cum}^{int}}(i,z') },
\end{equation}
where $\mathrm{N_{ion,cum}^{esc}}(z)$ and $\mathrm{N_{ion,cum}^{int}}(z)$ represent the cumulative number of escaping and intrinsic ionizing photons produced by a galaxy, summed over all of its progenitors over its entire assembly history. 

\begin{figure*}
\center{\includegraphics[scale=0.35]{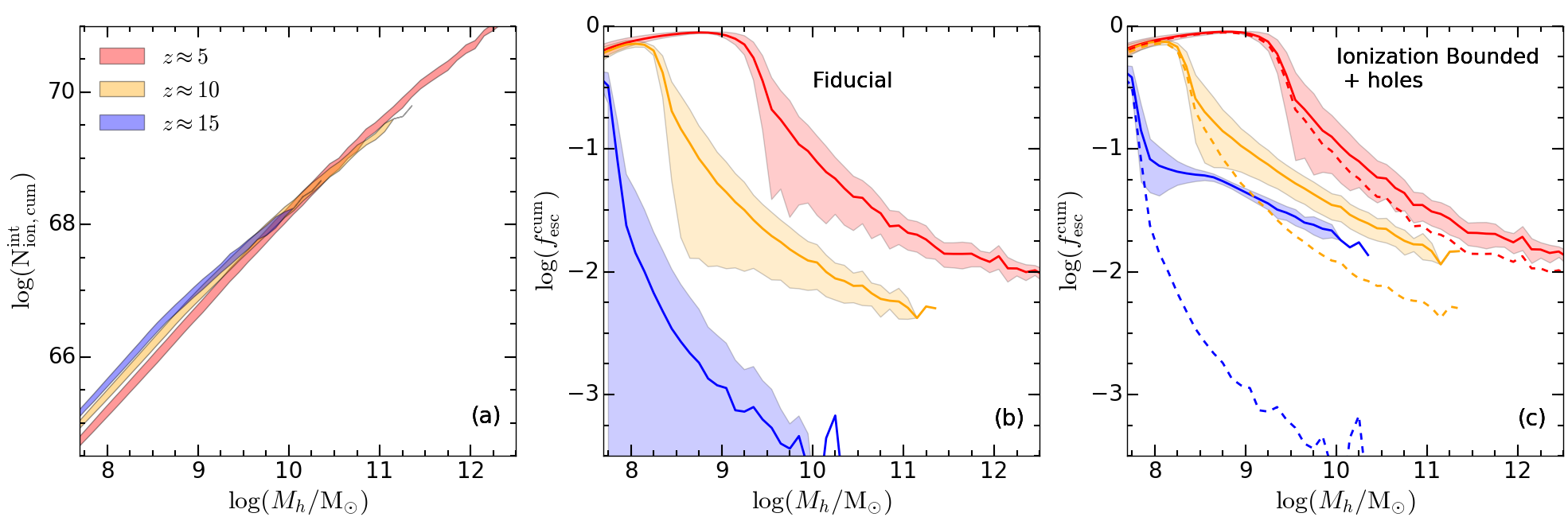}}
\caption{\small Cumulative number of intrinsically produced ionizing photons as a function of halo mass in panel (a);  cumulative escape fraction as a function of halo mass for the ionization bounded model in panel (b); and panel (c) shows the results for for the ionization bounded + holes model. All results are shown for $z\sim 5,10$ and 15. The shaded regions indicate $1-\sigma$ uncertainties. For comparison, we show the means from the ionization bounded model at $z\sim 5$ and 15 in the right panel.} \label{Nioncum}
\end{figure*}

As shown in panel (a) of Fig. \ref{Nioncum}, firstly, the cumulative intrinsic production rate of ionizing photons scales with the halo mass at all $z \sim 5-15$ and follows the stellar mass-halo mass relation as might be expected. Secondly, while the $\mathrm{N_{ion,cum}^{int}}-M_h$ slope is essentially independent of redshift, its normalization increases with increasing redshift by a factor of about $2.8$ between $z\sim 15$ and 5. This is a consequence of the star formation efficiency increasing with redshift for a given halo mass; we note that this relation shows a slight shallowing of its slope for $M_h \gsim 10^{9.2}\msun$ halos at $z \sim 15$ which is possibly driven by the low numbers of such halos that have assembled.  

We then discuss $\fescum$ for the {\it fiducial} model in panel (b) of Fig. \ref{Nioncum}. For all redshifts we find the same characteristic trend in which $\fescum$ declines with halo mass above the critical mass for ionization bounded leakage. For example, while a $10^{9}\mathrm{M_{\odot}}$ galaxy at $z\sim 5$ has $\fescum\sim 0.84$, a much more massive galaxy with $10^{11}\mathrm{M_{\odot}}$ has a value of $\sim 0.02$. 
This trend is driven by an increasing mass fraction of progenitors above the critical mass for leakage not contributing to $\mathrm{N_{ion, cum}^{esc}}$. This is also responsible for the strong impact of redshift on $\fescum$. As an example, at a fixed mass of $10^{9}\mathrm{M_{\odot}}$, we find $\fescum\sim 0.84$ by $z\sim 5$, which is reduced to $\sim 0.04~ (10^{-3})$ at $z\sim 10~ (15)$. This shows that for this particular mass, galaxies become orders of magnitude less efficient over their entire lifetime in injecting their intrinsically produced ionizing photons into the IGM.  We also encounter variations of up to two orders of magnitude in $\fescum$ at fixed halo mass above the critical mass for leakage, as partly shown by the $1-\sigma$ areas. This emphasises the impact of the variations in the underlying mass assembly. 
 
Finally, we discuss the results for the ionization bounded model with holes as shown in panel (c) of the same figure. We find that leakage in this model becomes increasingly important with increasing redshift: while at $z\sim 5$ both $\fesc$ models are fairly similar, by $z\sim 15$ the effective escape fraction is a factor of $\sim 38$ higher in the ionization bounded model with holes compared to the {\it fiducial} model for a $M_{h}\sim 10^{9}\mathrm{M_{\odot}}$ halo. This shows that over the entire lifetime of galaxies, the fraction of escaping ionizing photons is increasingly dominated by LyC leakage through SNII channels with increasing redshifts.  This model also shows a reduced scatter with respect to the {\it fiducial} model, since $\fesc > 0$ for any galaxy when leakage through channels is considered.

\section{Conclusions and discussion}
\label{conclusionsection}

In this work, we have combined the semi-analytical framework {\sc Delphi} for high-$z$ galaxy formation, with an analytic model to estimate the escape fraction of ionizing photons. The evolving ionization front within the gas distribution of each galaxy, and the release of SNII energy form the basis of this model. Leakage occurs either; when the ionization front reaches the virial radius (ionization bounded); or through a combination of ionization bounded leakage and additional leakage through gas-free channels created by SNII explosions.  

The key aim of this work is to understand the dependence of $\fesc$ on galaxy properties, its co-evolution with galaxy assembly, and the galaxy population driving reionization. Our main findings are:

$\mathrm{(i)}$  In the ionization bounded scenario, we find the central gas density to be primarily constraining $f_{\mathrm{esc}}$. A consequence of galaxies becoming denser at higher redshifts, is a shift in the leaking population from $M_{h}\lsim 10^{9.5}\mathrm{M_{\odot}}$ at $z\sim 5$ to $M_{h}\lsim 10^{7.8}\mathrm{M_{\odot}}$ by $z\sim 15$. 
Galaxies above this mass range are too dense to become fully ionized, hence leakage through gas-free channels plays a key role. 

$\mathrm{(ii)}$
We quantify the time-evolving leakage for a given assembly history of a galaxy in terms of $\effesc$. We find the co-evolution of the effective escape fraction with the assembly history of galaxies to be mass-dependent.
In the low mass $\mathrm{log}(M_{h}\mathrm{[M_{\odot}])}= 8.5-9.0$ bin, we see an increasing effective escape fraction with decreasing redshift.  
In the massive $\mathrm{log}(M_{h}\mathrm{[M_{\odot}])}= 11.0-11.5$ bin on the contrary, we find the opposite trend with a strong steepening in the slope at which $\effesc$ declines at $z\lsim 10$. 

$\mathrm{(iii)}$
While low mass $M_{h}=10^{7.7-9}\mathrm{M_{\odot}}$ galaxies spend $\gsim 90 \% $ of their lifetime in the $\effesc >0.50$ regime, massive $M_{h}=10^{12-13}\mathrm{M_{\odot}}$ galaxies spend roughly $80\%$ of their lifetime in the $\effesc<0.05$ regime. We find the variability to be driven by variations in the underlying assembly histories. 
As an example, in the narrow $M_{h}=10^{9.5-9.75}\mathrm{M_{\odot}}$ mass bin we find the fractional lifetime spent at $\effesc>0.50$ to range between 0 and 97 $\%$. As for the redshift dependence, for a fixed halo mass the lifetime fraction spent in a given $\effesc$ regime is declining with increasing redshift.

$\mathrm{(iv)}$
In our ionization bounded model reionization starts at $z\sim 16$ and is complete at $z=5.67$. When including leakage through holes, reionization ends roughly 50Myr earlier at $= 5.91$ and starts as early as the appearance of the first galaxies around $z \sim 30$, with a similar reionization history. Low mass $M_{h}< 10^{8.0}\mathrm{M_{\odot}}$ galaxies only contribute about $10\%$ to the ionizing background at $z\sim 5$, while at $z\sim 10$ this is $67\%$.
Galaxies purely leaking through channels contribute $6$ $(13)\%$ at $z\sim 5$ (15).

$\mathrm{(v)}$
Lastly, we defined the cumulative escape fraction, expressing the escaped fraction of all ionizing photons ever produced by a galaxy. Regardless of redshift, for both leaking models, we find $\fescum$ to decline with halo mass at masses above which the critical central density for ionization bounded leakage is reached. 
 
We end by stating a number of simplifications assumed in our model: (i) while we use a single physically motivated gas density profile, the specific $f_{\mathrm{esc}}$ trend with intrinsic galaxy properties sensitively depends on the particular gas density profile. Any physically motivated gas density profile however has to simultaneously match the emissivity and $\tau_{\mathrm{es}}$, reducing the freedom of choice of the particular density profiles; (ii) neglecting reionization feedback in our model, most likely leads to an over-prediction of the contribution of low mass galaxies to the EoR. While this is most likely the case at redshifts close to the end EoR, at higher redshifts this should not be significant; (iii) while we have only considered single stars, non-standard stellar populations such as binaries \citep[e.g][]{eldridge2017} could enhance the escape fraction and the reionization history of the Universe, which we will explore in a forthcoming paper; (iv) the shape of the gas density profiles is always assumed to be static, regardless of mergers or SN feedback processes; (v) relaxing our assumption of instantaneous clearing of gas channels might reduce the impact of leakage through holes; (vi) reducing (enhancing) our gas temperature threshold of $2\times 10^{4}$K would lower (enhance) the critical mass for ionization bounded leakage. 
When for example assuming $T_{\mathrm{gas}}=T_{\mathrm{vir}}$, it was already evident from Fig. \ref{cenden} that the central gas densities are significantly reduced for galaxies with $T_{\mathrm{vir}}>2\times 10^{4}\mathrm{K}$. For the ionization bounded case, this leads to $f_{\mathrm{esc}}\sim 0.99$ for $M_{h}\gsim 10^{9.5}\mathrm{M_{\odot}}$ at $z\sim5$. As early as $z\sim 15$, galaxies with high escape fractions extend up to  $M_{h}\sim 10^{10}\mathrm{M_{\odot}}$. This leads to an early completion of reionization by $z\sim8.5$, hence overestimating the electron scattering optical depth by $\sim 2.6\sigma$.

\section*{Acknowledgments} 
The authors express their appreciation to the referee for the insightful comments.
J. Bremer and P. Dayal acknowledge support from the European Research Council's starting grant ERC StG-717001 (``DELPHI"). P. Dayal also acknowledges support from the NWO grant 016.VIDI.189.162 (``ODIN") and the European Commission's and University of Groningen's CO-FUND Rosalind Franklin program. We thank Maxime Trebitsch for his useful input and insightful discussions.

\section*{Data Availability}
Data generated in this research will be shared on reasonable request to the corresponding author.

\bibliographystyle{mnras}
\bibliography{fesc}


\end{document}